**Phase Separation in Giant Planets: Inhomogeneous Evolution of Saturn**

Jonathan J. Fortney[1], William B. Hubbard[1]

[1]Lunar and Planetary Laboratory, The University of Arizona, Tucson, Arizona 85721-0092

E-mail: jfortney@lpl.arizona.edu

## ABSTRACT

We present the first models of Jupiter and Saturn to couple their evolution to both a radiative-atmosphere grid and to high-pressure phase diagrams of hydrogen with helium and other admixtures. We find that prior calculated phase diagrams in which Saturn's interior reaches a region of predicted helium immiscibility do not allow enough energy release to prolong Saturn's cooling to its known age and effective temperature. We explore modifications to published phase diagrams that would lead to greater energy release, and propose a modified H-He phase diagram that is physically reasonable, leads to the correct extension of Saturn's cooling, and predicts an atmospheric helium mass fraction $Y_{atmos} = 0.185$, in agreement with recent estimates. We also explore the possibility of internal separation of elements heavier than helium, and find that, alternatively, such separation could prolong Saturn's cooling to its known age and effective temperature under a realistic phase diagram and heavy element abundance (in which case Saturn's $Y_{atmos}$ would be solar but heavier elements would be depleted). In none of these scenarios does Jupiter's interior evolve to any region of helium or heavy-element immiscibility: Jupiter evolves homogeneously to the present day. We discuss the implications of our calculations for Saturn's primordial core mass.

## I. INTRODUCTION

The interiors of Jupiter and Saturn, extrasolar giant planets (EGPs), and brown dwarfs (BDs) are all described by similar physics: these bodies are mainly composed of liquid metallic hydrogen, and their interior energy transport mainly occurs through efficient convection (Burrows *et al.* 2001; Hubbard *et al.* 2002), leading to largely isentropic interiors. Jupiter and Saturn, whose radius, mass, luminosity, and age are known precisely, can serve as calibrators of thermal-history calculations for the entire class of objects. They can provide a test of the adequacy of the diverse physical models, including interior thermodynamics, heat transport mechanisms, and model-atmosphere grid, entering into the general thermal-history theory for EGPs and BDs. However, at very low effective temperatures (~ 100 K), the corresponding interior temperatures may become low enough for phase separation of abundant interior components to occur, and this effect must be quantitatively evaluated before Jupiter and Saturn can be used as calibrators. The purpose of this paper is to provide a quantitative assessment of inhomogeneous evolution in Jupiter and Saturn.

### A. Simplified Evolution Theory

A thermal-history calculation for an isolated nonrotating giant planet of mass $M$, radius $a$, and specified composition (atomic abundances relative to hydrogen, relative to solar abundances, an array symbolically denoted as $X$), yields relations of the form

$$L \equiv 4\pi\sigma\, a^2 T_{eff}^4 = L(M, t, X) \ ,$$
$$a = a(M, t, X) \ ,$$

(1)

where $L$ is the planet's luminosity, $\sigma$ is the Stefan-Boltzmann constant, $T_{eff}$ is the planet's effective temperature, and $t$ is the planet's age (time since accretion of its hydrogen envelope). Under the assumption of homogeneous evolution, i.e., that $X(r)$ = constant (where $r$ is the radius of a mass shell inside the planet) and $S(r)$ = constant (where $S$ is the entropy per unit mass of the deep interior), expressions (1) can be derived with the help of a grid of model atmospheres. The grid is obtained by choosing independent variables $T_{eff}$ and $g$ (atmosphere's surface gravity), integrating the atmospheric structure inward to a depth where it is fully convective and essentially isentropic, and then calculating $S$ at depth:

$$S = S(T_{eff}, g, X) \ ,$$

(2)

where the surface gravity is given by

$$g = GM/a^2 \ .$$

(3)

In addition to inhomogeneous evolution (phase separation), a number of other effects, leading to modifications of relations (1) and (2), must first be assessed for a description of the evolution of Jupiter and Saturn. These effects are as follows.

### B. Irradiation

Power absorbed from the Sun is a significant component of the total luminosity of both Jupiter and Saturn. Indeed, for EGP "roasters" such as 51 Peg B and HD209458 B, absorbed power from the parent star must contribute the preponderance of the planet's luminosity. Most properly, we should include irradiation in the grid of model atmospheres. Thus, a given model atmosphere of prescribed $g$, $T_{eff}$, and $X$ should include a self-consistent treatment of the absorption and scattering of solar photons at each level, and be integrated to the depth where the atmosphere stays convective and $S$ becomes constant. Although such self-consistent calculations are in progress and will be reported in future papers, here we make a traditional approximation that allows a treatment of irradiation within the framework of a grid of model atmospheres for isolated objects. The approximation consists of assuming that stellar photons scattered by the atmosphere do not cause its thermal structure to deviate from that of an isolated object, and that the remaining stellar photons are absorbed (thermalized) within the object's deep convective deep layers where $S$ is constant. In this approximation, the parameter $T_{eff}$ represents the effective temperature of all thermal photons radiated by the object into space; relation (2) for isolated objects is then used in unmodified form. However, if we equate $L$ to the intrinsic luminosity of the planet, i.e. to the total power radiated into space derived from the planet's interior, eq. (1) must be modified to

$$L \equiv 4\pi\sigma\, a^2 (T_{eff}^4 - T_{eq}^4) = L(M, t, X) \ ,$$
(4)

where $T_{eq}$ is the effective temperature that the planet would have if its $L = 0$, i.e., no intrinsic luminosity. We derive $T_{eq}$ from the Bond albedo, $A$, according to

$$4\pi\sigma\, a^2 T_{eq}^4 = (1-A)\pi\, a^2 L_S / 4\pi\, R^2 \ ,$$
(5)

where $L_S$ is the solar luminosity and $R$ is the Sun-planet distance. As described in Hubbard et al. (1999), we use an expression for $L_S = L_S(t)$ from solar evolution models. However, we do not yet have an adequate model for $A(t)$. Thus we use the measured present-day $A$ for both planets for all $t$. This simplification is relatively minor since Jupiter and Saturn have spent the past ~3 Gyr with $T_{eff}$s < 20% warmer than their current values. As will be discussed later, uncertainties in the atmosphere grid and the H/He EOS have larger effects on evolution calculations.

Heat flow parameters used in this paper for the present-day Jupiter and Saturn are given in Table 1; values are derived from the review paper of Conrath *et al.* (1989).

**Table 1**
**Heat flow parameters**

| planet | $T_{eff}$ (K) | $A$ | $T_{eq}$ (K) |
|--------|---------------|-----|--------------|
| Jupiter | 124.4 ± 0.3 | 0.343 ± 0.043 | 110.1 ± 1.3 |
| Saturn | 95.0 ± 0.4 | 0.344 ± 0.040 | 81.3 ± 1.0 |

### C. Non-adiabaticity in Interior?

This correction arises, in principle, because in the outer (mostly ideal-gas) layers of the planet, the Rosseland-mean photon opacity at thermal wavelengths may fall below the value necessary to cause convective instability (Guillot *et al.* 1994). However, as discussed by Guillot *et al.* (2002), the presence of alkali elements in solar proportions may suffice to supply enough opacity to restore convective instability in the region from 1 to10 kilobars. On the other hand, a variant of the dense-hydrogen equation of state proposed by Ross (1998) has the interesting property, at pressures ~ 1 megabar,

$$\left( \frac{\partial P}{\partial T} \right)_\rho < 0, \tag{6}$$

where $\rho$ is the mass density. In a region where Eq. (6) is true, the material is always stable to convection regardless of the temperature gradient. There is a controversy about whether Ross' model has theoretical justification (see Hubbard *et al.* 2002, Guillot *et al.* 2002), and we do not explore its consequences in the present paper. Instead, we assume that all interior regions of the planet in which $X$ is constant have piecewise continuous specific entropy. At interfaces such as a phase boundary (e.g. between molecular and metallic hydrogen), we take $T$ and $P$ to be continuous with a corresponding jump in $S$.

### D. Rotation

Both Jupiter and Saturn are rapid rotators, and are consequently nonspherical. As the planet evolves and contracts, the rotation rate and axial moment of inertia changes. Since the spin angular momentum is conserved to good approximation, some of the heat lost from the interior goes into increasing the planet's spin kinetic energy, which must be deducted from the luminosity. However, as shown by Hubbard (1970), this effect produces only a small correction to the thermal evolution age $t$. Therefore we ignore the effect of rotation in the energy balance calculation.

Similarly, for purposes of the present paper, we replace the rotationally distorted Jupiter or Saturn with an equivalent spherical planet of the same mass and surface area. As described below, the equivalent spherical planet at $t = 4.56$ Gyr (present) is fitted to constraints on the present-day Jupiter and Saturn (mass, radius, and axial moment of inertia).

### E. Correction for Chemical Inhomogeneity in Interior

The present-day atmospheric composition, age, and intrinsic heat flow of the giant planets Jupiter and Saturn are, in principle, coupled. However, detailed quantitative estimates of the variation of atmospheric abundances and heat flow as a function of age $t$ have been lacking. The seminal papers of Stevenson and Salpeter (1977a, 1977b) predicted that both Jupiter and Saturn might have limited solubility of helium in the metallic-hydrogen fluid interior, leading to a depletion of the atmospheric He abundance and an extension of the cooling age beyond the value for chemically homogeneous evolution. Stevenson (1975) made the first quantitative prediction of the two-component phase diagram of a hydrogen-helium plasma at multi-megabar pressures. First-generation evolutionary models for Jupiter and Saturn (Pollack et al. 1977; Hubbard 1977, Grossman et al. 1980) showed that the observed heat flow from Jupiter was consistent with homogeneous evolution, but that chemically-homogeneous models for Saturn evolved too fast, typically passing through the present-day heat flow value one to two gigayears before present. Initial Voyager results for the atmospheric helium abundances in both Jupiter and Saturn showed a marked depletion of helium relative to solar ratio, suggesting that inhomogeneous evolution was indeed important for both bodies (Conrath *et al.* 1984). In a post-Voyager review paper Hubbard and Stevenson (1984) presented a rough analysis of the expected extension of Saturn's cooling age due to He sedimentation. The analysis indicated rough consistency with the planet's age and the Voyager atmospheric He abundance.

The observational and theoretical situation has since become less clear-cut. The shortfall in Saturn's cooling age for homogeneous evolution seems robust (e.g., Hubbard *et al.* 2002), but the evidence for pronounced atmospheric He depletion has eroded. Table 2 summarizes available data for atmospheric abundances in Jupiter and Saturn. We do not present in Table 2 the Voyager result for Jupiter's He mass fraction, $Y = 0.18 \pm 0.04$ (Gautier *et al.* 1981), as it is fully superseded by results from the two He-abundance experiments on the Galileo Entry Probe, the HAD (Von Zahn *et al.* 2000) and the NMS (Niemann *et al.* 2000). The solar $Y$ value in Table 2 represents the solar atmospheric value, apparently slightly reduced from the primordial solar system value, $Y_{primordial} = 0.27$, due to diffusion in the Sun (Bahcall *et al.* 1995). The HAD and NMS values for Jupiter's $Y$ are in good agreement, with the more-precise HAD result about $6\sigma$ lower than $Y_{primordial}$ and about $3\sigma$ lower than the present-day solar atmospheric $Y$. Thus, Jupiter's atmospheric He depletion is less marked than the Voyager result, but still significant.

**Table 2**

**Detected Atmospheric Elemental Abundances in Jupiter and Saturn**

| element | solar mass fraction | Galileo NMS mass fraction | Galileo HAD mass fraction | Saturn-Voyager mass fraction | Saturn revised mass fraction |
|---------|---------------------|---------------------------|---------------------------|------------------------------|------------------------------|
| H | 0.736 | 0.742 | 0.742 | 0.92 | 0.76 |
| He | 0.249 | $0.231 \pm 0.04$ | $0.231 \pm 0.006$ | $0.06 \pm 0.05$ | $0.215 \pm 0.035$ |
| C | 0.0029 | $0.009 \pm 0.002$ | | ~0.01 | |
| N | 0.00085 | $\leq 0.012$ | | ~0.004 | |
| O | 0.0057 | $\leq 0.0035$ | | | |
| Ne | 0.0018 | $\leq 0.0002$ | | | |
| P | 0.00001 | $\leq 0.00007$ | | ~0.00003 | |
| S | 0.00050 | $0.00091 \pm 0.00006$ | | | |
| Ar | 0.00007 | $\leq 0.00015$ | | | |
| "Z" | 0.015 | 0.027 | | ~0.02 | |

The row labeled "Z" in Table 2 represents, for the Sun, the expected mass fraction of all elements beyond H and He ("metals" in astrophysical parlance). For the Jupiter and Saturn, "Z" represents the estimated mass fraction of *detected* elements beyond H and He.

The error bars for the Voyager result for Jupiter, $Y = 0.18 \pm 0.04$, and the HAD result, $Y = 0.231 \pm 0.006$, do not quite overlap, and there has been concern that a systematic problem of some sort, possibly in the radio-occultation profile for the atmospheric temperature $T$ as a function of pressure $P$, might affect the Voyager $Y$ values for both Jupiter and Saturn. Conrath and Gautier (2000) revisited the derivation of Saturn's $Y$ from Voyager infrared radiometry. They fitted Saturn's infrared flux with a self-consistent $T$-$P$ profile, rather than making use of the radio-occultation $T$-$P$ profile, and derived the value in the last column of Table 2, finding $Y = 0.215 \pm 0.035$, instead of the much lower Voyager result, $Y = 0.06 \pm 0.05$.

According to the Galileo NMS results, Jupiter's atmospheric $Z$ is at least twice the solar value. Since Jupiter's carbon abundance is about three times solar, Jupiter's deep oxygen abundance could be enhanced by a similar factor (Owen, *et al.* 1999). In an alternate scenario (Gautier, *et al.* 2001a, 2001b), Jupiter's deep oxygen abundance might be more than nine times solar. It is possible that the NMS oxygen abundance is affected by jovian meteorology (Showman and Ingersoll 1998), since a gradual increase in the abundance was measured up to the maximum depth reached by the probe. Note that the solar value for the oxygen abundance has recently been revised downward (Allende Prieto *et al.* 2001). As a result, the Galileo NMS's upper limit for Jupiter's deep oxygen abundance increases to 60% of solar (previously it was 35% of solar, based on Grevesse and Sauval 1998). Assuming that the Galileo NMS abundance for oxygen *is* affected by meteorology, and that the true deep oxygen abundance in Jupiter, like carbon, is actually three times solar, $Z$ in Jupiter's envelope might be as large as $\sim 0.05$. Based on a detailed parametric study of Jupiter's overall $Z$ abundance using Galileo probe results, Guillot, Gautier, and Hubbard (1997) concluded that for Jupiter, $0.04 < Z < 0.14$. In this paper, we do not carry out a detailed study of Jupiter or Saturn's $Z$. Rather, in Sections II and III.A-D we parameterize our spherical models by a spatially and temporally constant value of $Z$ in the hydrogen-helium layers of the planet. This value of $Z$, which is determined by using a water equation of state to represent all of the "ices", will be called $Z_{ice}$ in the following. The value of $Z_{ice}$ is inferred by adjusting it to give the correct mean radius and axial moment of inertia of the planet at the present epoch. As has been noted in previous studies (e.g., Hubbard and Stevenson 1984, Guillot *et al.* 1997), static models of Jupiter and Saturn give values of $Z_{ice}$ comparable in magnitude to $Y$ (in contrast to solar composition). Thus the possibility of significant inhomogeneous evolution involving separation of the ice component, instead of helium, cannot be ruled out. In Section III.E we schematically explore the possibility of redistribution of $Z_{ice}$ as Saturn evolves.

Hubbard et al (1999) made the first quantitative attempt to evaluate the inhomogeneous evolution of Jupiter and Saturn, using the expression (cf. Eq. 4)

$$L(M, t, X) = -M \int_0^1 dm T \frac{\partial S}{\partial t} , \qquad (7)$$

where the dimensionless mass-shell variable $m$ is defined by

$$m = \frac{1}{M} \int_0^r 4\pi r'^2 dr' \rho(r') . \qquad (8)$$

Equation (7) gives the heat extracted from the planet's interior per unit time.  The equation is valid for homogeneous evolution, where the isentropic interior evolves to progressively lower-entropy adiabats.  When inhomogeneous evolution occurs, the equation is still valid, but more heat can be extracted because denser matter (e.g., more He-rich matter) has lower entropy per unit mass $S$ at given $T$ and $P$.  Thus, keeping the interior fixed on a $T(P)$ relation corresponding to the surface condition (Eq. 2), heat can be extracted as the outer layers (at lower $T$) lose a dense component, causing $S$ to rise in these layers, while the deeper layers (at higher $T$) increase in density, causing S to decrease in these layers.

## II. PROPOSED PHASE DIAGRAMS

### A. The "Standard" Theory of H-He Mixtures

Figure 1 shows a high-pressure phase diagram for hydrogen (Hubbard et al. 2002, Guillot et al. 2002), with calculated $T$-$P$ relations in present-day Jupiter and Saturn's hydrogen-rich layers superimposed (heavy solid lines).  The dashed extensions of these curves to the upper right show estimates of the course of Jupiter and Saturn's interiors in their deepest non-hydrogen-rich cores.

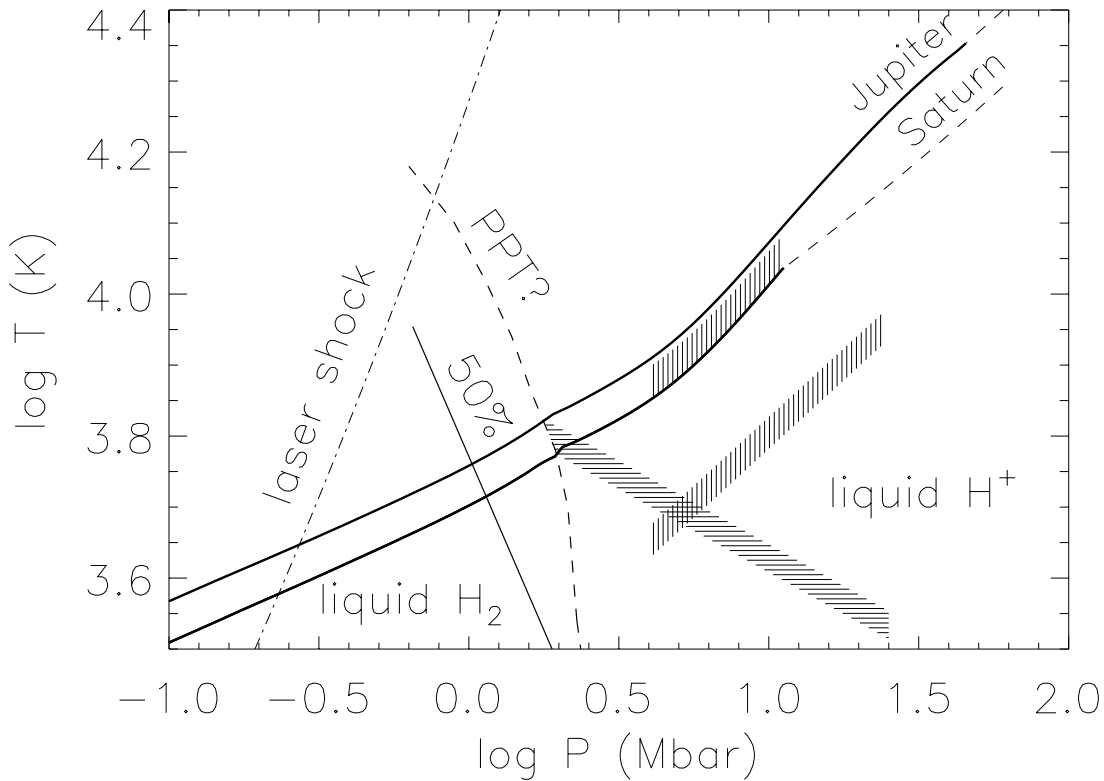

Figure 1.  Temperature-pressure plot of the interiors of Jupiter and Saturn at t = 4.56 Gyr, superimposed on a hydrogen phase diagram (see text for details).  The upper boundary of the horizontally-hatched region shows the minimum temperature at which He is fully miscible in metallic hydrogen with a mass fraction $Y = 0.27$, while the lower boundary shows the minimum temperature corresponding to $Y = 0.21$, according to H-D theory.  The lower vertically-hatched region shows the same He miscibility limits according to Pfaffenzeller et al. (1995), while the upper vertically-hatched region shows the modification to the Pfaffenzeller et al. theory that gives a realistic prolongation of Saturn's age (see Section III).

In the *T-P* range shown in Figure 1, there is only one relevant phase boundary for pure hydrogen: the boundary between liquid molecular hydrogen ($H_2$) and liquid metallic hydrogen ($H^+$). The dashed curve marked "PPT?" shows the (controversial) boundary between these two phases, the so-called Plasma Phase Transition or PPT calculated by Saumon et al. (1995; SCVH). The solid line labeled "50%" shows an alternative theory (Ross 1998) for the boundary between the mainly molecular and mainly metallic states of hydrogen. According to Ross, there is a smooth transformation rather than an abrupt first-order phase transition between the two states; the curve indicates where half of the H atoms are in $H_2$ molecules and half are unbound atoms. The dot-dashed line on the left side of Figure 1 shows the computed shock-compression trajectory of the laser implosion experiments of Collins *et al.* (1998). The hatched areas show possible regions of limited miscibility of He in $H^+$, to be discussed below. There are no experimental data yet available concerning the miscibility of He in $H^+$.

The theory of Stevenson (1975) and of Hubbard and DeWitt (1985; HDW) give equivalent results for the miscibility of He in metallic hydrogen. Both theories apply perturbation theory to a model of fully pressure-ionized He and H. For convenience, we use the HDW theory rather than the Stevenson theory to calculate miscibility limits representative of either theory. Using a computer-coded version of the HDW theory, we calculate the Gibbs free energy $G$ per atom of a H-He mixture, using

$$G = \frac{E - TS + (P/\rho)}{N} \ , \tag{9}$$

where $N$ is the total number of H and He atoms per gram, $E$ is the internal energy per gram, and $1/\rho$ is the volume per gram. Letting $N_H$ be the number of H atoms per gram and $N_{He}$ the number of He atoms per gram, we define the He mole fraction to be

$$x = \frac{N_{He}}{N} \ , \tag{10}$$

and we define the usual Gibbs free energy of mixing (per atom), $\Delta G$, by

$$\Delta G = G - x G(x=1) - (1-x) G(x=0) \ . \tag{11}$$

It is important to note that the SCVH theory for a H-He mixture is produced by a linear superposition of the thermodynamics of pure hydrogen and pure helium (apart from the mixing entropy), and thus SCVH theory does *not* include a description of H-He immiscibility. However, HDW theory does include the essential parts of Stevenson's (1975) theory for describing separation into two liquid phases. For our purposes here, we treat the Stevenson theory and the HDW theory as interchangeable, equivalent to the "standard" theory of H-He mixtures in the liquid-metallic H regime. Figure 2 shows some representative calculations from HDW theory for a pressure of 5 Mbar, indicating separation into two coexisting phases with differing values of *x*, according to the double-tangent construction.

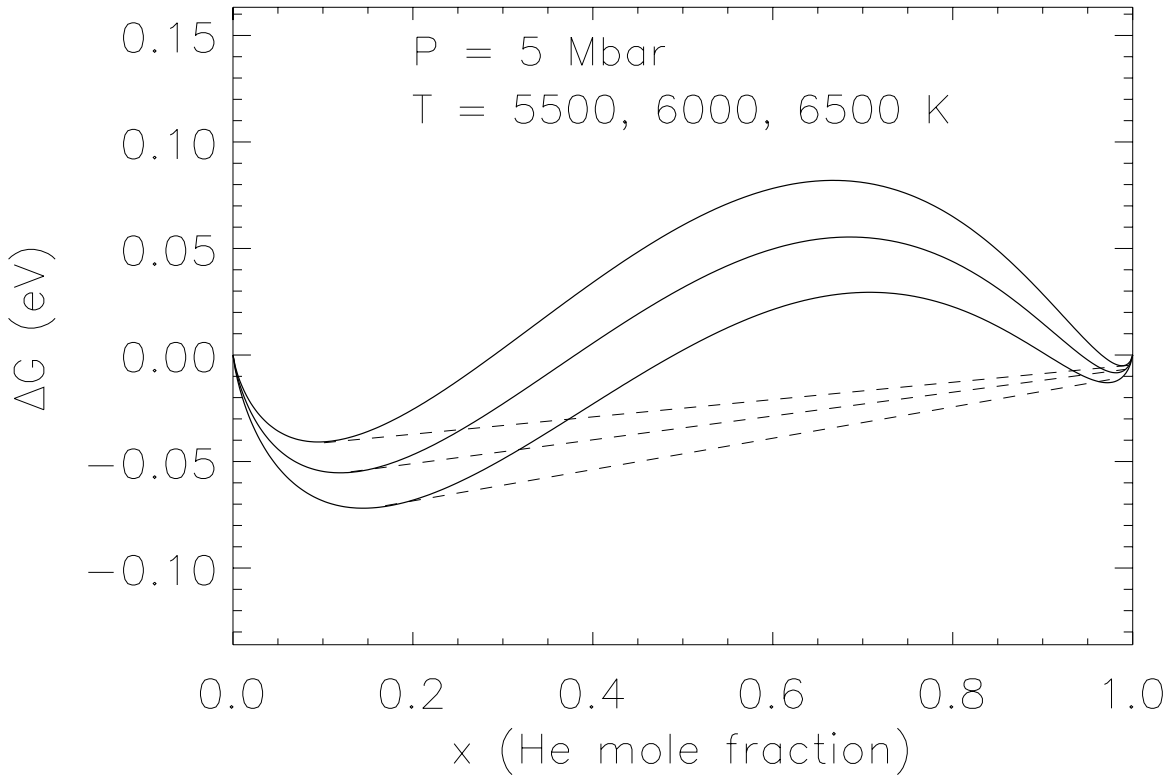

Figure 2. Sample results for the Gibbs energy of mixing $\Delta G$ of H and He, according to HDW theory. Separation into He-poor and He-rich liquid phases is shown by means of the double-tangent construction (dashed lines; lowest dashed line corresponds to the highest temperature).

## B. Parameterization of Theories of H-He Mixtures

For the purpose of conveniently incorporating phase separation in calculations of evolving planetary models, we require convenient parametric representations for fairly complex equilibrium curves. Figure 3 shows the full equilibrium curve (plusses) for a H-He mixture at 5 Mbar, computed using HDW theory. The vertical straight line shows the value of $x$ corresponding to primordial nebular composition, $Y = 0.27$. As the diagram indicates, when a H-He mixture of this composition cools to a temperature below the equilibrium curve delineated by the plusses (about 5000 K at 5 Mbar), essentially pure He droplets (with $x \approx 1$) will separate out. Thus, we make the approximation that it is only necessary, for purposes of evolutionary models, to describe the phase equilibrium curve in the vicinity of an initial primordial composition, $x \approx 0.08$, with the further approximation that the coexisting phase has $x \approx 1$. Using HDW theory, we carried out a sequence of calculations over the pressure range appropriate to liquid-metallic hydrogen in the Jovian (and Saturnian) interior, and parameterized the results according to

$$\log_{10} T = c_1 \log_{10} P + c_2 \log_{10} x + c_3 \ , \tag{12}$$

where the best fit to HDW theory gives $c_1 = -0.234$, $c_2 = 0.315$, and $c_3 = 4.215$, for $T$ in K and $P$ in Mbar.

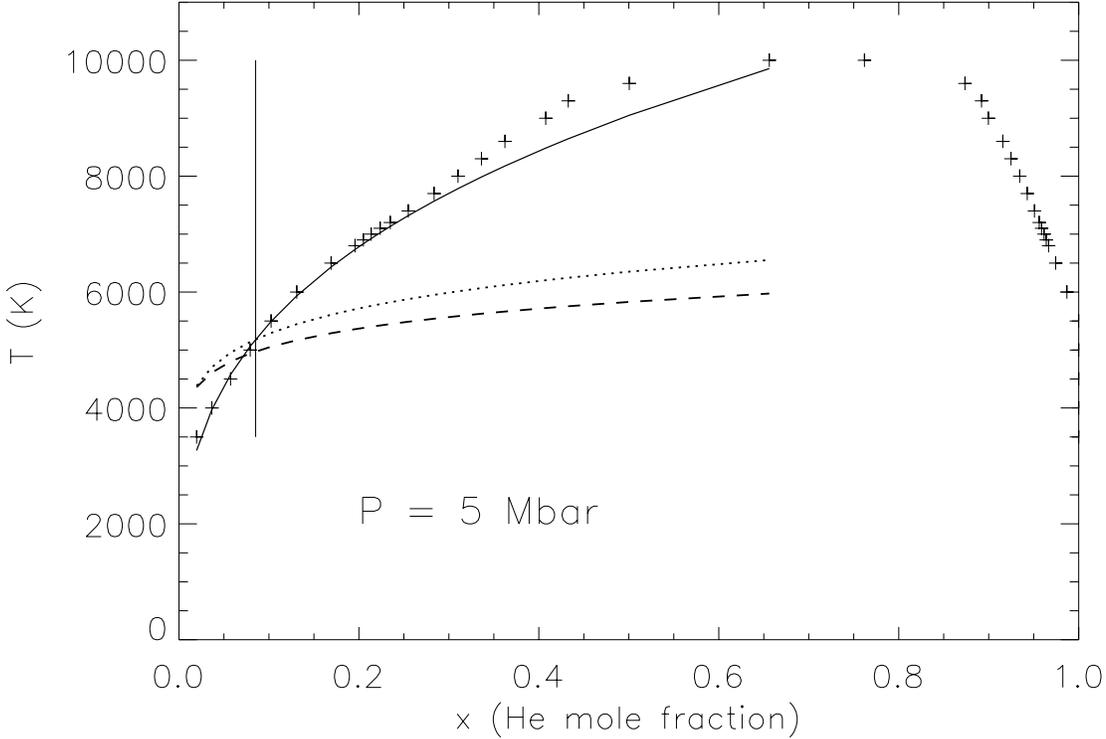

Figure 3. Plusses show the equilibrium curve for H-He mixtures, according to HDW theory. The vertical line shows primordial nebular composition; the solid curve shows the fit of Eq. (12) to the HDW theory. The two dashed curves show possible (unphysical) modifications to the HDW theory that would give a correct Saturn age (discussed in Section III, and there labeled Trials 2 and 3).

As we see in Section III, the unmodified HDW theory does not give the correct Saturn age. In order to modify the theory, as well as incorporate alternative theories, we need some guidance as to the physical significance of the constants in Eq. (12) or its equivalent.

A common approximation (see Stevenson 1979 and Pfaffenzeller et al. 1995) for the saturation value of $x$ is written

$$x = \exp\left(B - A/k_B T\right),\tag{13}$$

where $B$ is a dimensionless constant, $k_B$ is Boltzmann's constant, and $A$ is a constant with units of energy. We can estimate the constants $B$ and $A$ by assuming that for $x \ll 1$, and at $T \sim 5000$ K, the internal energy and pressure can be replaced with their zero-temperature values and that the entropy is dominated by the ideal entropy of mixing. Thus we approximate Eq. (9) with

$$G \approx \frac{E_0 + (P_0/\rho)}{N} - T\frac{S_{mix}}{N} = \frac{H_0}{N} - T\frac{S_{mix}}{N},\tag{14}$$

where $E_0$, $P_0$, and $H_0$ are respectively the internal energy, pressure, and enthalpy per gram at zero temperature, and $S_{mix}$ is the ideal entropy of mixing of a liquid composed of H and He,

$$\frac{S_{mix}}{N} = k_B x \ln x + k_B (1-x) \ln (1-x) \ . \tag{15}$$

If we make the approximation $x \ll 1$ and minimize $G$ approximation (14) yields

$$B \approx 0, \ A \approx \frac{1}{N} \left( \frac{\partial \Delta H_0}{\partial x} \right)_{x=0} \ , \tag{16}$$

such that $A$ can be interpreted as the increase in enthalpy upon addition of a helium atom to pure liquid-metallic hydrogen. Here $\Delta H_o$ is the zero-temperature enthalpy of mixing, defined analogously to Eq. (11),

$$\Delta H_0 = H_0 - x H_0 (x=1) - (1-x) H_0 (x=0) \ . \tag{17}$$

Fitting Eq. (13) to explicit HDW calculations at $P = 5$ Mbar and $x = 0.08$, we obtain $B = 0.64$, $A = 1.37$ eV. The value of $B$ is not precisely zero because of limitations of approximation (14) and because $x$ is not extremely small. The value of $A$ agrees well with Stevenson's (1979) estimate of $\sim 1.6$ eV. Figure 4 shows a comparison of Eq. (12) and other parametric representations of the form of Eq. (13) with the H-He phase curve of HDW.

Note that $A$ determines the rate of change of x with respect to the interior temperature at a fixed pressure; hence the larger the value of $A$, the larger the contribution of phase-separation to the planet's luminosity. For fixed $A$, the value of $B$ determines the temperature at which phase separation commences: the larger $B$, the higher the temperature and the earlier the epoch.

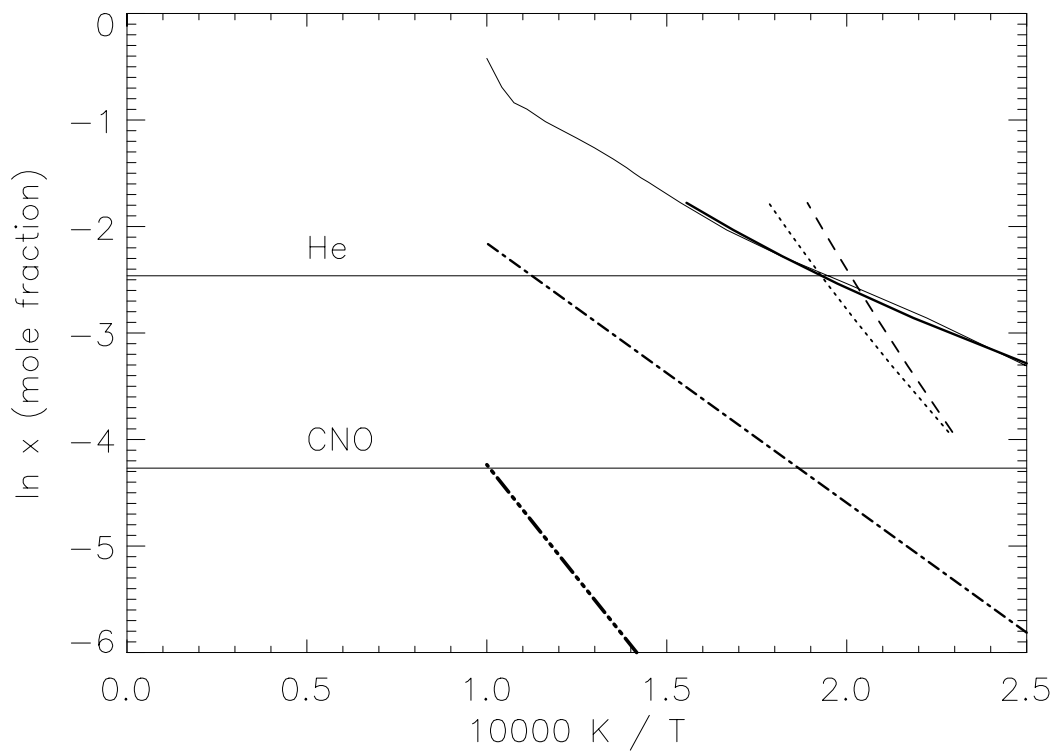

Figure 4. An alternate representation of Figure 3, to show the adequacy of approximation (13). Light solid curve shows the maximum He content of liquid-metallic hydrogen at P = 5 Mbar, according to HDW theory. Upper horizontal line shows primordial He abundance. Heavy solid curve shows Eq. (12) evaluated at 5 Mbar. The two dashed curves show (probably unphysical) modifications to the HDW theory that would give a correct Saturn age (Trials 2 and 3, discussed in Section III). The upper dot-dashed curve shows a modification to the theory of Pfaffenzeller et al. (1995), in which He immiscibility *increases* with pressure, that would give a correct Saturn age (Section III.D). The lower horizontal line and lower dot-dashed curve illustrate the point that unmixing of a less-abundant element, such as C, N, or O, with a larger value of *A*, could also give a correct Saturn age (Section III.E).

### C. Phase diagram of Pfaffenzeller *et al.*

The theories of Stevenson (1975) and HDW assume full pressure ionization of hydrogen and helium. Pfaffenzeller *et al.* (1995) use a molecular dynamics method to calculate the region of He immiscibility, taking into account a consistent treatment of the electronic structure around the nuclei. They find that the location and shape of the He separation region is substantially different for a *partially ionized* plasma. For the region of interest in their calculation, 4 to 24 Mbar (pressures greater than 4 Mbar are likely safely within the region of $H^+$), they find a critical demixing line with a $T(P)$ slope opposite to that of Stevenson and HDW (this would correspond to a positive value of $c_1$ in Eq. 12). The region for the onset of helium separation in a solar composition mixture runs nearly parallel to planetary internal adiabats. However, the critical temperature for He separation is nearly a factor of 2 lower than values estimated for Saturn's interior, meaning Saturn could not reach this region in a Hubble time.

The Pfaffenzeller *et al.* theory can be represented by analytic expressions of the form of Eq. (13). Independently fitting Eq. (13) to the Pfaffenzeller *et al.* results, we find that *A* is an increasing function of pressure and can be described by

$$A = 1.0 \text{ eV} \left( \frac{P}{4} \right)^{0.387},$$ (18)

for *P* in Mbar, while by hypothesis *B* = 0 at all pressures. Thus, Pfaffenzeller *et al.*'s theory gives *A* = 1.09 eV at a pressure of 5 Mbar. The $T(P)$ relation implied by Eqs. (13) and (18), for *x* fixed at the primordial nebular value 0.08, lies essentially parallel to Jupiter and Saturn adiabats on a log T vs. log P plot. This $T(P)$ relation lies *below* Jupiter and Saturn adiabats (see Figure 1), implying that neither planet would unmix helium. However, in Section III.D we show that fairly minor modifications, probably well within the range of uncertainty of the Pfaffenzeller *et al.* theory, can be made to *A* and *B*, leading to a Saturn model that enters the region of He separation and then evolves to the current age and current intrinsic heat flow.

## III. EVOLUTIONARY MODELS

### A. Homogeneous Evolution

We first create evolutionary models for Jupiter and Saturn for which the planet's interior remains homogeneous during its entire cooling history. Our equation of state is the "interpolated" equation of state of hydrogen and helium of SCVH. We use the non-gray isolated-object grid of model atmospheres of Burrows *et al.* (1997). With these in place we can calculate the time step between successive models in an evolutionary sequence. Eq. (7) can be rewritten as

$$\delta t = -\frac{M}{L} \int_0^1 T\, \delta S\, dm, \qquad (19)$$

where δt is the time step, and the other variables have the same meanings as defined earlier.

For the equation of state for our rocky core we use the ANEOS equation of state of olivine. This may perhaps be overly dense compared to the planets' actual cores, which are thought to be composed of an unknown mixture of rock and ices. For the heavy element enhancement in the hydrogen envelope we use the $H_2O$ equation of state of ANEOS, and assume the volume-additivity rule. The first five columns of Table 3 give the planet and the observed quantities that constrain the final model. The last three columns give inferred parameters for homogeneous evolution.

**Table 3**
**Planetary Parameters**

| Planet | Known Age (Gyr) | $T_{eff}$ (K) | Mean $a$ (km) | $C/Ma^2$ | Model Age (Gyr) | Model $Z_{ice}$ | Model $M_{core}$ ($M_\oplus$) |
|---|---|---|---|---|---|---|---|
| Jupiter | 4.56 | 124.4 | 69235 | 0.264 | 4.7 | 0.059 | 10 |
| Saturn | 4.56 | 95.0 | 57433 | 0.220 | 2.1 | 0.030 | 21 |

As shown in Figure 5, with our homogeneous models, Jupiter reaches $T_{eff}$ = 124.4 K in 4.7 Gyr, while Saturn reaches $T_{eff}$ = 95.0 K in 2.1 Gyr. Theses results are very similar to those found by many other authors, including Hubbard *et al.* (1999), which lends confidence that the simplifications made in our evolution models are acceptable.

Uncertainties in the atmospheric boundary condition and the H/He EOS make up the bulk of the uncertainties in modeling the evolution of giant planets. The Burrows *et al.* (1997) model atmosphere grid is the best atmospheric boundary condition currently available for modeling the evolution of giant planets and brown dwarfs. The grid is nongray and utilizes the predicted abundances of atoms and molecules in chemical equilibrium. It should be noted that this grid is for solar metallicity, whereas the atmospheres of Jupiter and Saturn appear to have 3 and ~ 5 metallicity enhancements in their atmospheres. The increased opacity of a metal enhanced atmosphere should tend to slow a planet's evolution, but not nearly enough to make up Saturn's shortfall. Knowledge of the EOS of hydrogen and helium has been a long-standing problem in the modeling if giant planets and leads to the greatest uncertainty in evolutionary calculations. Recent shock experiments (Collins *et al.* 1998) on hydrogen indicate deficiencies in the SCVH "interpolated" EOS, but even these experiments are the subject of controversy (Knudson, *et al.*, 2001). The SCVH EOS is the best currently available for astrophysical applications. Uncertainties in the EOS could lead to errors of up to ~0.4 Gyr in evolution calculations for Jupiter and Saturn.

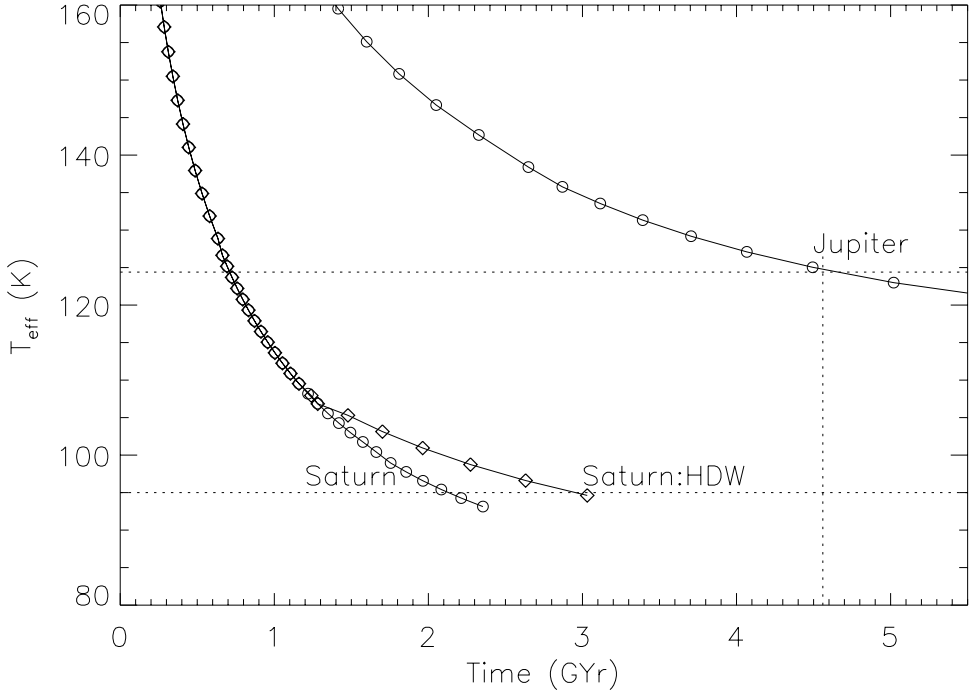

Figure 5. Homogeneous evolution of Jupiter and Saturn, along with inhomogeneous evolution of Saturn using the standard theory of helium immiscibility of HDW. This increase in the cooling time due to the HDW phase diagram is actually an upper limit, as discussed in the text. Marked in dotted lines are the age of the solar system (4.56 Gyr) and the current effective temperatures of Jupiter and Saturn, 124.4 K and 95.0 K, respectively. The inhomogeneous Saturn model's $Y_{atmos}$ falls to 0.215 at an effective temperature 95.0 K.

## B. Inhomogeneous Evolution

Inhomogeneous evolution adds a complicating factor to the evolution picture. When the initially protosolar ratio helium/hydrogen mixture encounters a region of pressure and temperature is which helium is immiscible, the helium separates into two phases. As noted earlier, one phase is essentially pure helium. Enough helium separates out such that the remaining helium is miscible in the hydrogen. The helium that separates out will coalesce via diffusion to form droplets. These droplets will fall to deeper layers in the planet under the influence of gravity, despite convection (Stevenson and Salpeter 1977b). The droplets then redissolve at higher pressures in the H⁺ when they leave the immiscibility region. Since the interior is fully convective up to the planet's visible atmosphere, helium from lower-pressure layers nearer to the planet's surface will continuously mix down into the immiscible region. We assume that this happens instantaneously (relative to evolutionary timescales). Layers at pressures lower than that of the immiscible region become helium-depleted, and layers at pressures higher than that of the immiscible region become helium-enriched. An observable consequence is that the planet's atmosphere becomes depleted in helium relative to the protosolar ratio.

As discussed in detail by Stevenson and Salpeter (1977b), the He concentration gradient has important effects on the temperature gradient in the immiscibility region. It is well known that composition gradients tend to inhibit convection, such that a steeper temperature gradient is needed to maintain convective instability. In addition, the temperature at the bottom in the immiscibility region is the starting point for the helium-enriched adiabat in the homogeneous helium-enriched regions of the deep H[+] interior. Stevenson and Salpeter (1977b) found that the temperature gradient in the immiscibility region matches a condition for overstability, and estimated the thickness of the immiscibility region to be $10^3$-$10^4$ km, which is ~1.5-15% of the planet's radius, a fairly large range.

Deriving the actual temperature gradient is non-trivial because it involves interplay with the phase diagram. If, at a given pressure, the temperature is increased, this allows a greater $Y$ to be miscible at that pressure. But this then increases the composition gradient, which leads to a steeper temperature gradient, and again allows a greater $Y$ at this pressure. Here we find the two limits to the temperature gradient in the immiscibility region, and investigate the effect on the evolution of Saturn.

The first limit would be that the composition gradient does not affect the temperature gradient. This would be the smallest temperature gradient possible and would lead to the He gradient region taking up the largest amount of the planet's radius. We find this lower limit by essentially setting $Y$ for the depleted (outer) layer, and then calculating $Y$ in the gradient region and homogeneous enriched (inner) regions of the planet, with the constraints that the mass of the helium in the planet must be conserved, and $Y$ in the gradient region given by the maximum allowed by the HDW phase diagram. The adiabatic temperature gradients of the homogeneous inner and outer regions are joined in the center of the immiscibility region.

We attempted to find an upper limit to the temperature gradient by setting the gradient equal to the limit for convective stability. This is the boundary between overstability and instability. The temperature gradient could be steeper than this condition, if the overstable modes are inefficient (Stevenson and Salpeter 1977b). However, as described later, this condition did lead us to the maximum possible temperature gradient.

Our limit is derived as follows. We have

$$\frac{dP}{d\rho} = \left(\frac{\partial P}{\partial \rho}\right)_{T,x} + \left(\frac{\partial P}{\partial T}\right)_{\rho,x} \frac{dT}{d\rho} + \left(\frac{\partial P}{\partial x}\right)_{T,\rho} \frac{dx}{d\rho}, \tag{20}$$

and since

$$\left(\frac{\partial P}{\partial \rho}\right)_{S,x} = \left(\frac{\partial P}{\partial \rho}\right)_{T,x} + \frac{T}{\rho^2 C_V}\left(\frac{\partial P}{\partial T}\right)_{\rho,x}^2, \tag{21}$$

after manipulation, the temperature gradient at the convective stability limit can be written

$$\frac{dT}{dP} = \frac{T}{\rho^2 C_V}\frac{d\rho}{dP} + \frac{dx}{dP}\left(\frac{\partial T}{\partial x}\right)_{P,\rho}, \tag{22}$$

where the first term is the standard adiabatic temperature gradient and the second term has two factors due to the composition gradient. The first factor, $dx/dP$, is the actual composition gradient and the second factor, $\partial T/\partial x$, must be calculated from the H/He EOS. In our situation both factors are always positive, with $\partial T/\partial x$ being on the order of a few $\times 10^5$ K.

In the absence of an *a priori* constraint on the thickness of the immiscibility region, this stability limit leads to a vanishing thickness for the inhomogeneous region and a very steep temperature gradient because of the large positive value of $\partial T/\partial x$. For example, if we consider a case where the planet's atmospheric helium mass fraction $Y_{atmos}$ has dropped to 0.215, overall conservation of He in the planet requires that the interior helium mass fraction (below the inhomogeneous layer) $Y_{interior} = 0.35$. The phase diagram then requires that there be a temperature jump equal to 1300 K as we reach the bottom of the inhomogeneous layer. Deeper interior layers continue on this warmer, He-enriched adiabat.

For our lower limit for the temperature gradient, the entire planet cools as a region of the interior falls through the region of helium immiscibility. For the upper limit of the temperature gradient, as the helium-depleted ($H_2$) exterior cools, the helium-enriched ($H^+$) interior becomes increasingly *warmer*, due to the increasingly large step in temperature in the gradient region. These two possibilities lead to very different evolutionary histories. For a cooling history with the upper limit of the temperature gradient, the deep interior is much warmer (and hence has a higher entropy) than for the case with the minimum temperature gradient. If the interior is kept at a high entropy state, little energy is released, so the helium separation does little to affect the planet's cooling. In the second case (the minimum temperature gradient), the entropy drops rapidly in the interior, and much energy is released, which allows the planet to remain at a high $T_{eff}$, and hence, retards its cooling. Stevenson and Salpeter's (1977b) analysis found that the temperature *does* increase in the deep helium-enriched interior, so the actual temperature gradient is likely to be closer to the upper limit than the lower limit.

The size of the immiscibility region will grow as the planet cools. This is due to two reasons. First, the predicted immiscibility region is roughly triangular on a log $P$ – log $T$ plot. More importantly, the size of the immiscibility region (in $P$-$T$ phase space) is larger for a greater helium mass fraction—immiscibility will occur at a higher temperature. Since the maximum ratio of helium to hydrogen that is miscible is a function of pressure and temperature, in the immiscibility region of Saturn's interior just deeper than hydrogen's molecular-metallic transition, a gradient in helium mass fraction is established. At a given mass shell in the gradient region, the helium mass fraction is the maximum allowed by the HDW phase diagram.

With our inhomogeneous evolution code we are able to test whether the HDW phase diagram leads to enough helium separation to prolong Saturn's cooling to its current known effective temperature and age. Figure 5 shows the results of homogeneous evolutionary models of Jupiter and Saturn, as well as the evolution of Saturn with helium separation, using the phase diagram of HDW and PPT hydrogen transition phase line of SCVH. This transition pressure is ~ 1.9 Mbar in our region of interest. Both planets are assumed to have an initial helium mass fraction $Y = 0.27$. Jupiter reaches its current effective temperature and age just short of reaching the HDW helium immiscibility region, so helium separation has no effect on its evolution.

In Saturn, helium separation starts at $T_{eff} \sim 107$ K, and its atmospheric $Y$ falls to 0.215. The extension of cooling shown, $\sim 0.8$ Gyr, is for the *lower* limit for the temperature gradient in the immiscibility region. This is the maximum extension of cooling. The cooling curve for the *upper* limit for the temperature gradient (not shown) plots on top of the homogeneously evolving Saturn—the energy from helium separation goes into heating the deep interior, but does not change the planet's luminosity. This is the minimum temperature extension of cooling. Since the actual temperature gradient will be between these limits, the extension of cooling will be between 0 and 800 million years. Figures 6 and 7 show the change of $Y$ and $S$ during the helium-separation phase of the planet's evolution for the minimum helium immiscibility temperature gradient. Figures 8 and 9 show the same, but with the maximum temperature gradient. Apparent in the figures is the partitioning of helium toward deeper layers, at the expense of helium in the molecular region, as well as the drop in $S$ in the planet's helium-rich layers. The greater the decrease in $S$, the greater the energy release, and consequently, the greater effect helium separation has on the planet's cooling. In the maximum temperature gradient case, since the interior temperatures become warmer, $S$ falls only modestly. Since the maximum extension of Saturn's cooling only allows the planet to reach 3.0 Gyr at an effective temperature at 95.0 K, we conclude that if helium were the only species currently differentiating in Saturn, and this differentiation were Saturn's only additional energy source, the HDW phase diagram must be incorrect.

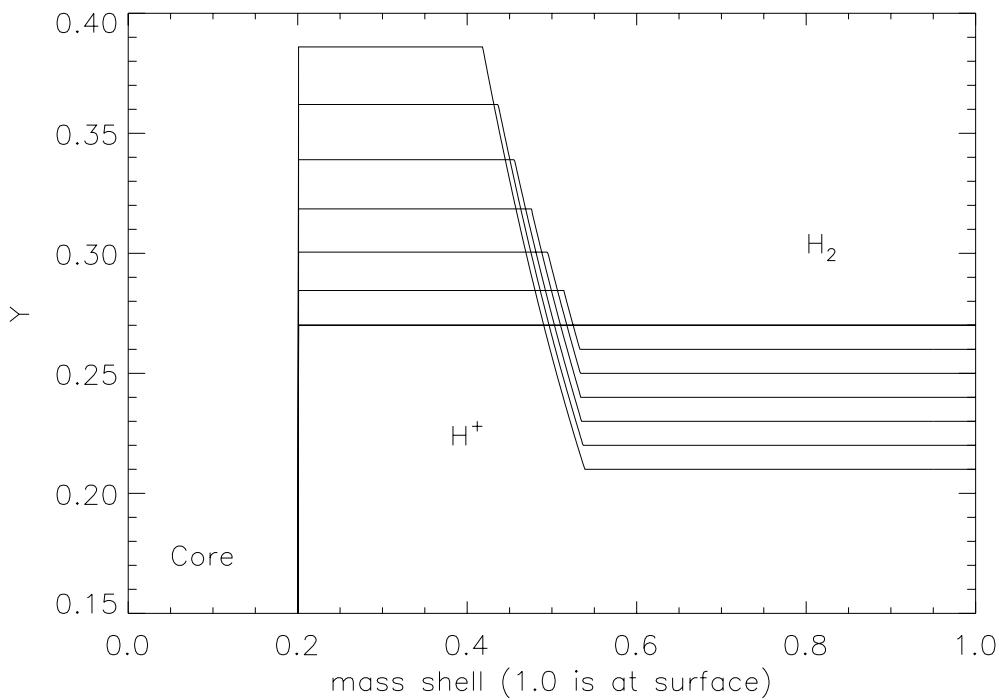

Figure 6. Saturn's interior helium mass fraction for the *minimum* temperature gradient limit as the planet evolves into the helium immiscibility region. Both Figures 6 and 7 show the last 9 Saturn models used to create the "Saturn HDW" cooling curve in Figure 5. However, here the first 3 models overlap because their interior maintains a constant $Y = 0.27$. Helium is lost from the $H_2$ region and gained in the $H^+$ region.

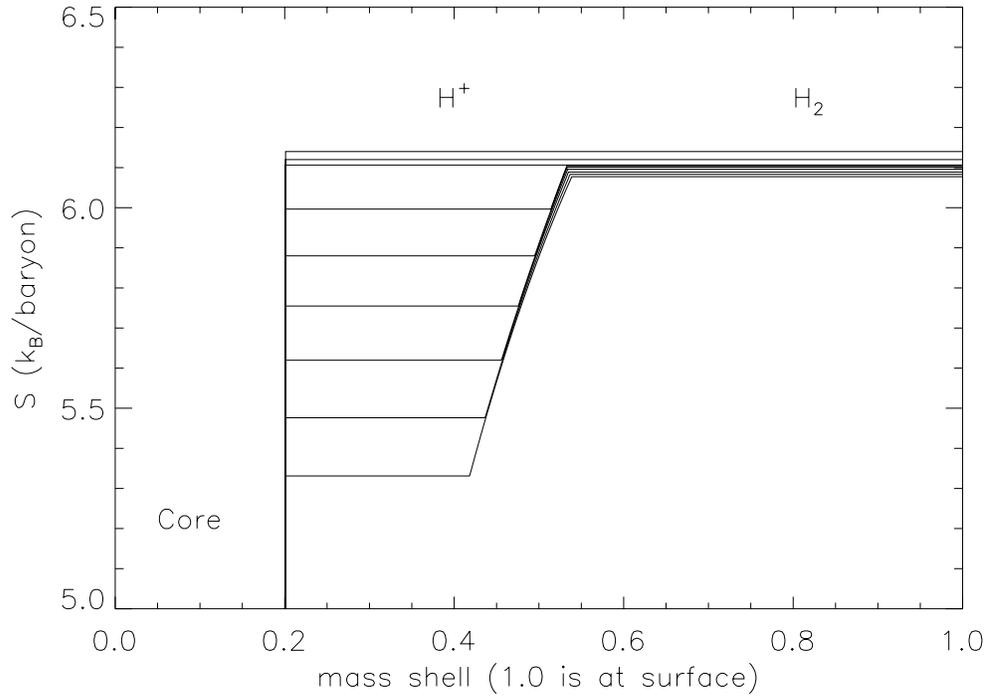

Figure 7. Saturn's interior entropy in the *minimum* temperature gradient limit as the planet evolves in the helium immiscibility region. All hydrogen rich layers are homogeneous and posses the same entropy until a portion of the planet starts to lose He. The He gradient region grows as the planet cools. The nine models shown are the same as in Figure 6. The progressive drop in $S$ of the $H_2$ region is stalled as $S$ drops significantly in the $H^+$ region. The transition from $H_2$ to $H^+$ remains nearly constant at a mass fraction of 0.53. The core, here the inner 20% of the planet's mass, is assumed to not take part in the planet's evolution.

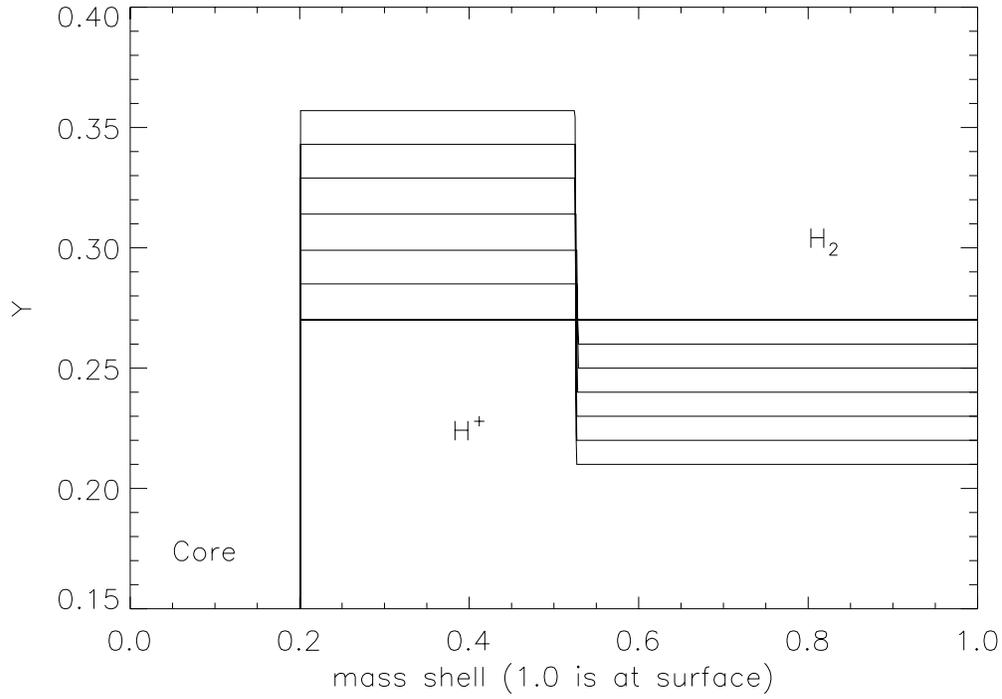

Figure 8. Similar to Figure 6, but here for the *maximum* temperature gradient limit as the planet evolves into the helium immiscibility region. Helium is lost from the $H_2$ region and gained in the $H^+$ region.

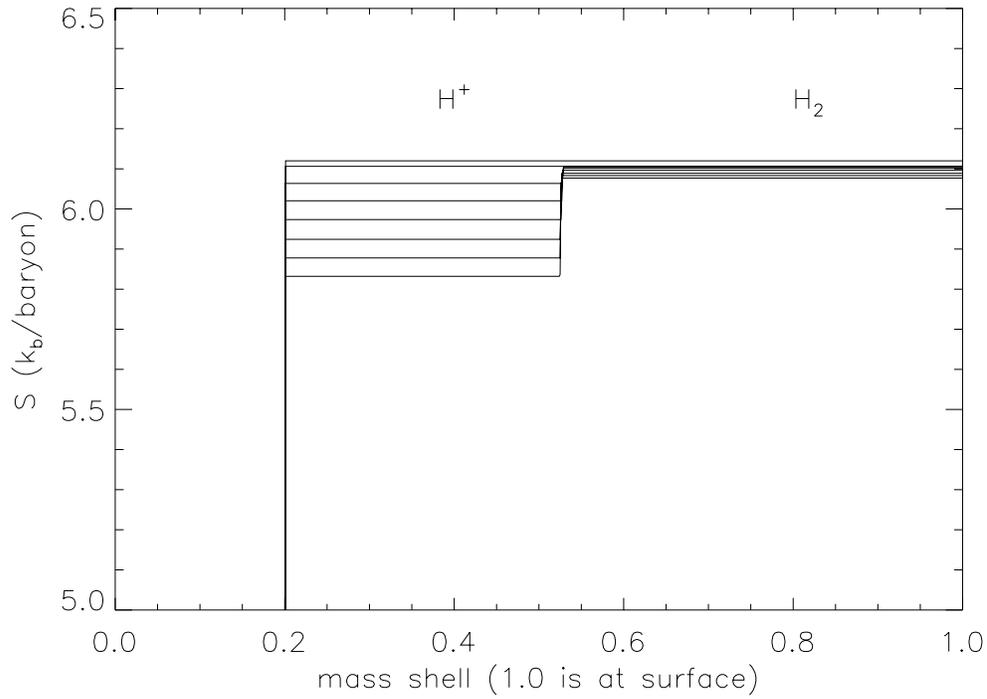

Figure 9. Similar to Figure 7, but here for the *maximum* temperature gradient limit as the planet evolves in the helium immiscibility region. All hydrogen rich layers are homogeneous and posses the same entropy until a portion of the planet starts to lose He.  The nine models shown are the same as in Figure 8.  The drop in S in the $H^+$ is relatively modest compared to Figure 7, because interior temperatures increase with increased $Y$, instead of decreasing.  Again, the transition from $H_2$ to $H^+$ remains nearly constant at a mass fraction of 0.53.

### C. Modified HDW Phase Diagrams

The constants $A$ and $B$ are fixed in HDW theory.  How much must these constants be modified to allow models of Saturn to reach the planet's known effective temperature at its current age of 4.56 Gyr?  And how physically reasonable is the resulting phase diagram?

Our two alternate phase diagrams are shown in Figure 3 (dashed curves).  Both are modifications of the HDW phase diagram, keeping the same value of $c_1$ in Eq. (12), but with a change in shape such that as the planet cools, and its adiabat falls into the unmixing region, the helium-poor phase has less helium than predicted by the HDW phase diagram, leading to more helium raining down for a given temperature decrement.  The first alternate phase diagram (designated as Trial 2 in Table 4 below) was required to have the same onset temperature as HDW for the start of He separation.  According to HDW theory combined with our model-atmosphere grid, helium separation begins in Saturn at $T_{eff}$ = 107 K.  Trial 2 supposes that the H-He phase diagram is similar to HDW theory in pressure dependence and temperature of onset of He separation, but that the actual temperature dependence of the effect is much steeper.  As a result, the helium mass fraction in the planet's $H_2$ region must fall to $Y = 0.10$ to allow enough heat release to prolong Saturn's age to ~ 4.56 Gyr at $T_{eff}$ = 95.0 K.  Figure 10 shows, on the $P$-$T$ plane, how HDW theory is modified for Trial 2.

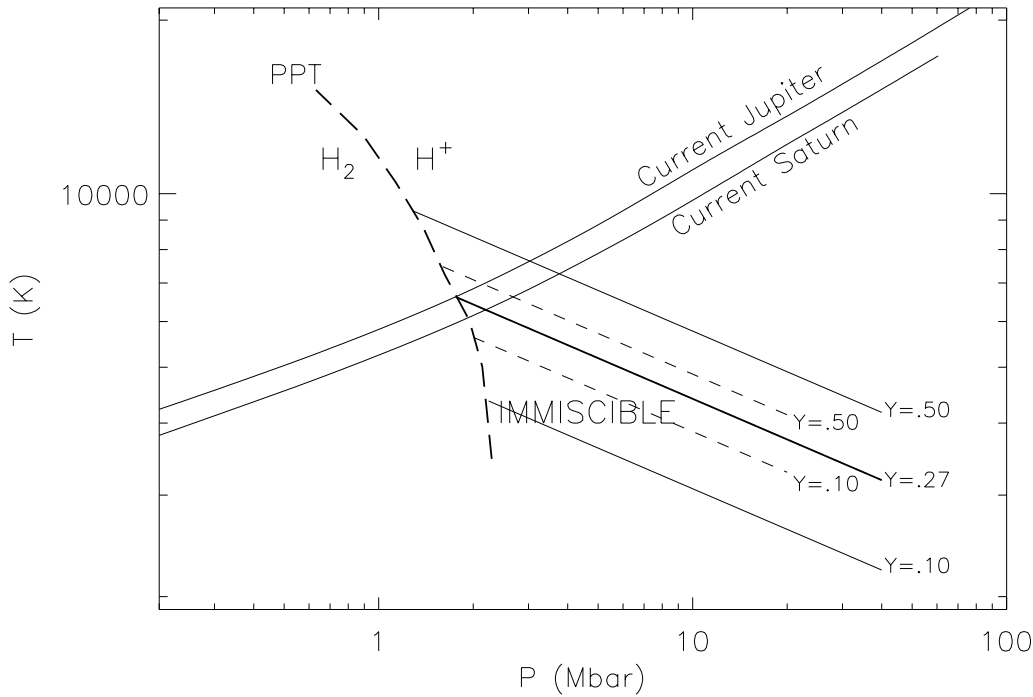

Figure 10.  The phase diagram of HDW with our first trial at modification (Trial 2 in Table 4).  Also shown are current Jupiter and Saturn adiabats, as well as the SCVH phase line (PPT), which we chose as our $H_2/H^+$ transition boundary.

Solid diagonal lines are the immiscibility boundaries from HDW theory. Shown are lines for helium mass fraction $Y$ = 0.10, 0.27 (protosolar), and 0.50. Dashed diagonal lines show the modification to the phase diagram. Unmixing at $Y$ = 0.27 remains in the same position, but all other unmixing lines are brought in closer, meaning more He separation will occur than for HDW, for the same drop in internal temperature. This modification results in an increase in both $A$ and $B$.

The second alternate phase diagram (Trial 3) allows the planet to cool longer before He separation begins. We suppose that Saturn does not reach the immiscibility region until $T_{eff}$ = 98.5 K, only 3.5 K above its current $T_{eff}$. This phase diagram differs even more from that predicted by HDW. Because the additional energy from unmixing is added later in Saturn's evolution, the age prolongation is greater for a given mass of He unmixing; the atmospheric He mass fraction falls to 0.13 at 4.56 Gyr. The evolution of the model Saturns in Trial 2 and 3 can be seen in Figure 11. It is important to note that the minimum temperature gradient was assumed in these modified HDW phase diagram trials, leading to the maximum possible prolongation of cooling.

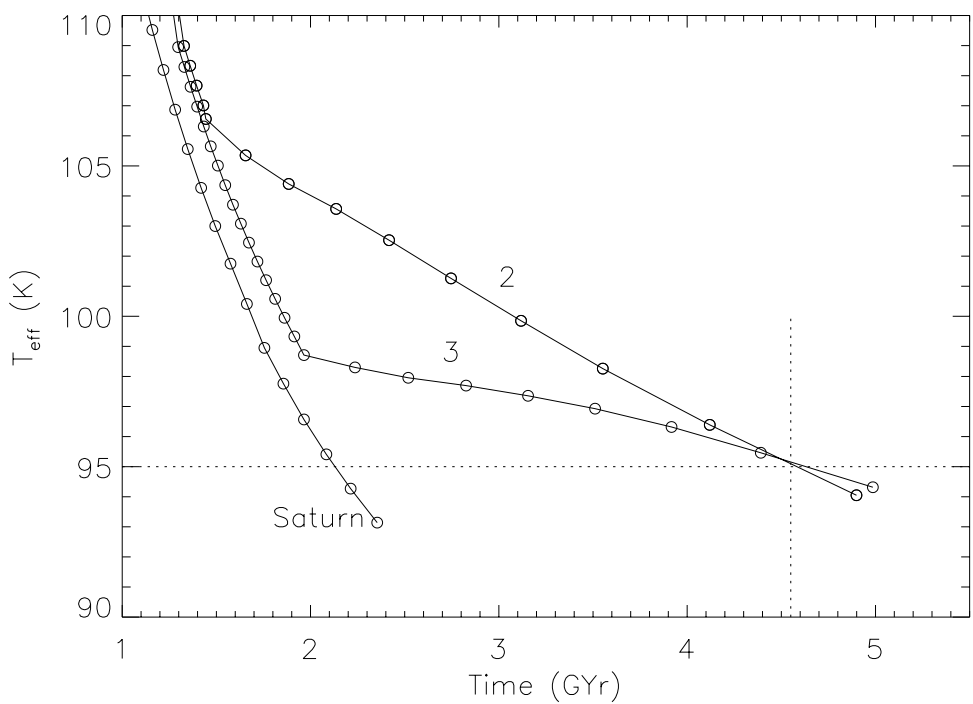

Figure 11. Cooling of Saturn after modifications to the HDW phase diagram. Homogeneous evolution and evolution of Trials 2 and 3, as described in Table 4, are shown. Both allow Saturn to reach ~ 4.56 Gyr at 95.0 K, although the minimum temperature gradient in the immiscibility region is used, as described in the text. This leads to the maximum prolongation of the planet's cooling. The evolutionary paths of the models before the onset of He separation are slightly different due to the differences in core masses and hydrogen-envelope heavy element enhancement between the models. The phase diagram for both trials is probably not physically realizable. Cooling curves for other pairs of Trials (4-5 and 6-7) look very similar to those of 2 and 3. For Trials 2-7, modifications to the value of $B$ are unrealistic.

But from their $B$ constants (see Table 4 below), it seems unlikely that either of these modified phase diagrams could be physically realizable. As discussed above, in the limit of small He mass fraction, the $B$ constant should be close to zero, while trials 2 and 3 have $B$ constants of 6.3 and 8.7 respectively. So while they allow Saturn to reach its current age, it is unlikely that these phase diagrams correspond to reality.

In addition, we constructed phase diagrams similar to those described, but with the $H^+/H_2$ transition pressure moved to 1.0 Mbar (below the PPT) or 3.0 Mbar (above the PPT). The evolution proceeded in the same fashion as described. As before, the $B$ constants were in general too large for the phase diagram to be physically plausible.

One item of note from this analysis is a general trend: the higher the transition pressure, the greater the mass fraction of helium left in the planet's molecular hydrogen region. This is due to the fact that the higher the $H_2/H^+$ transition pressure, the greater the total mass of the $H_2$ region relative to the $H^+$ region in the planet, so there is a larger reservoir of helium to be brought by convection into the immiscibility region. The helium mass fraction does not need to decrease as much in order to transport the mass of helium necessary to prolong the planet's cooling. It seems unlikely that a modified phase diagram in the style of HDW, in which helium separates out of a small region of the planet, but then redissolves into the $H^+$ below the immiscibility region without falling down to settle onto the core, can explain Saturn's current luminosity, if helium separation is the planet's only additional energy source.

### D. Modifications to the Pfaffenzeller *et al*. Phase Diagram

The most recent calculation of the location of the helium immiscibility region is that of Pfaffenzeller *et al.* (1995). For a solar composition mixture of helium and hydrogen, Saturn's interior does not enter the immiscibility region, as seen in Figure 1. Nevertheless, the predicted position of helium immiscibility is plausibly near the internal adiabat of Saturn, so we created a modified version of the Pfaffenzeller *et al.* phase diagram with a higher temperature for the onset of immiscibility. The modification scheme was similar to that of the modified HDW phase diagrams, in which the shape of the phase diagram was changed as shown in Figure 3. Again, we found the constants $A$ and $B$ by fitting Eq. (13) to the modified phase diagram. The shape of the immiscibility region is much different from that predicted by HDW, having a positive $c_1$. Immiscibility *increases* with increasing pressure and the upper boundary of the region runs nearly parallel to the interior adiabats.

In our modified phase diagram, we simplified the evolution calculations by slightly changing the slope of the top of the immiscibility region to be exactly parallel to the adiabats. As in our HDW trials, Jupiter still does not enter this region in the age of the solar system. As in the original Pfaffenzeller *et al.* phase diagram, the immiscibility region extends from 4 to 24 Mbar, affecting a large fraction of the liquid-metallic hydrogen region, and encompassing the pressure at the liquid metallic hydrogen / core boundary (~ 10 Mbar). Therefore, when helium separates from liquid metallic hydrogen it sinks towards the center of the planet and settles onto the planet's core, forming a pure helium layer. This is the same style of evolutionary models discussed in Hubbard *et al*. (1999). Helium is lost from *all* hydrogen-rich regions of the planet, due to the convective transport of helium amongst all layers, and this helium settles onto the core.

A small decrease in the helium mass fraction in the hydrogen can lead to a significant amount of evolutionary change in the planet. Starting helium separation at $T_{eff} = 107$ and 98.5 K respectively (Trials 8 and 9 in Table 4 below), as in previous models, these phase diagrams can lead to the necessary prolongation in evolution with only relatively modest decreases in $Y_{atmos}$, to 0.185 and 0.20 respectively. The latter reproduces the result of Hubbard *et al.* (1999): if helium separation started late in the planet's evolution (at $T_{eff} \sim 98.5$ K) and helium was lost from all hydrogen layers down to the core, a decrease in the atmospheric $Y$ to only 0.20 could prolong Saturn's cooling to reach $T_{eff} = 95.0$ K at 4.56 Gyr. Since there is no region where a gradient in composition exists, just a helium-poor and pure helium region, the planet remains fully convective and the adiabatic temperature gradient holds. Therefore, our predictions of Saturn's extension of cooling for this phase diagram are *not* upper limits, but are our actual predictions for the modified Pfaffenzeller, *et al.* phase diagrams.

Of the modified phase diagrams that lead to sufficient prolongation of Saturn's evolution, the Pfaffenzeller *et al.* (1995) style of phase diagram has the highest likelihood of being physically realized. Trial 8 increases the value of $A$ to 2.1 eV, somewhat higher than but similar to all previous calculations (Stevenson, HDW, Pfaffenzeller *et al.*), while $B$ is close to zero, as expected for the theory of binary mixtures at low concentration. Trial 8 leads to enough helium separation to sufficiently prolong Saturn's evolution, while leaving an atmospheric He mass fraction that falls within the error bars of the Conrath and Gautier (2000) analysis. Figure 12 shows the evolution of Saturn for Trials 8 and 9. Figure 13 shows the evolution of $S$ in the planet for Trial 8 after the onset of helium separation.

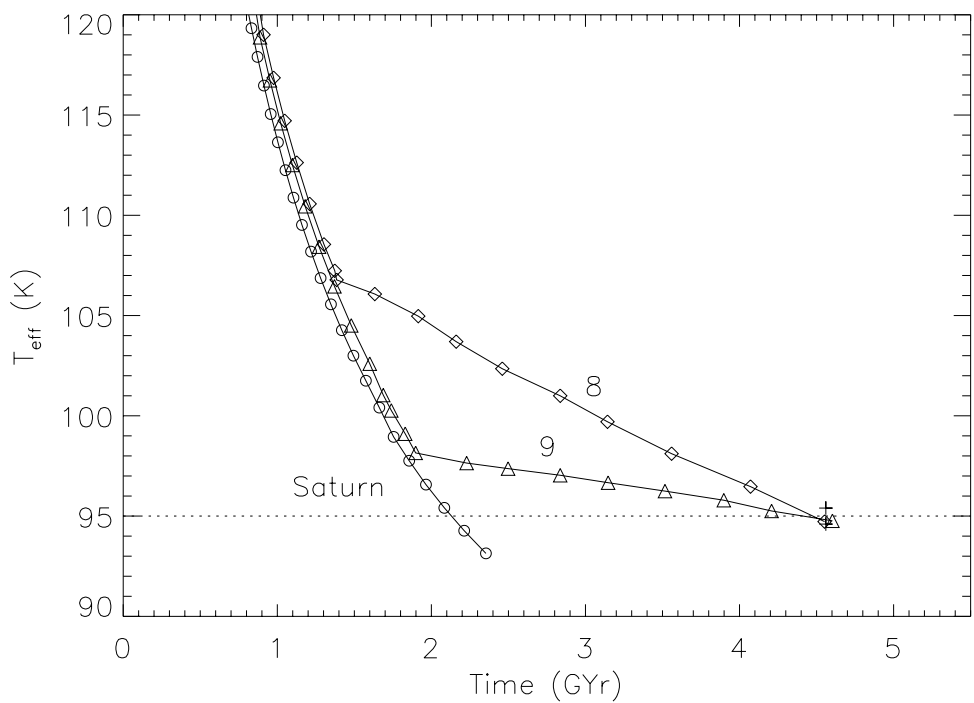

Figure 12. Cooling of Saturn after modifications to the Pfaffenzeller, *et al.* (1995) phase diagram. Homogeneous evolution and evolution of Trials 8 and 9, as described in Table 4, are shown. Both allow Saturn to reach ~ 4.56 Gyr at 95.0 K. These cooling curves are *not* upper limits, as they were for the HDW phase diagram. (See text.) The evolutionary paths of the models before the onset of He separation are slightly different due to the differences in core masses and hydrogen envelope heavy element enhancement between the models. The phase diagram for Trial 8 is probably physically realizable, but less so for Trial 9.

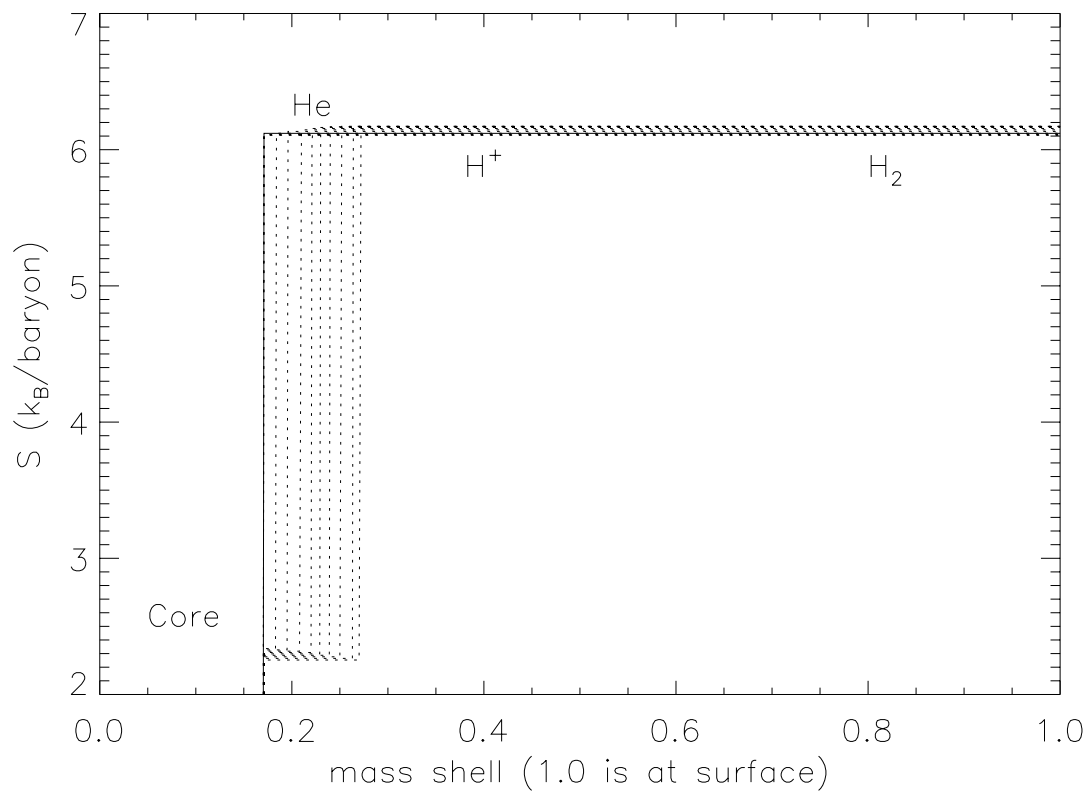

Figure 13. Evolution of Saturn's interior entropy distribution under the modified phase diagram of Trial 8. The pure helium layer on top of the core has much lower specific entropy than the hydrogen rich regions, leading to a release of energy. Helium rains down from all hydrogen rich regions, causing the pure helium region to grow as the planet cools.

### E. Separation of Heavier Elements

Helium may not the only element that has a limited solubility in hydrogen at the temperatures and pressures encountered in giant planets. Since the Galileo entry probe measurements have provided strong evidence that Jupiter has an abundance of heavy elements around 3 times solar, and Saturn has perhaps a larger enhancement, the question of whether separation of "metals" could have an appreciable effect on a planet's evolution naturally arises.

We created an evolutionary model for Saturn in which a heavy element or group of heavy elements separates from the hydrogen-helium mixture. In the planet's initial state the heavy elements are uniformly mixed throughout the envelope. We generically call this evolution "CNO element separation," and use the ANEOS $H_2O$ equation of state as described earlier. Our modeling of the CNO separation is not as detailed as that of He separation, since there are no computed phase diagrams that may shed light on the area of pressure-temperature space where these elements may become immiscible. In this trial, we created another Pfaffenzeller *et al.* style phase diagram, in which the "ices" are lost uniformly from the H/He envelope and form a pure layer on the core. For our CNO separation phase diagram we constrained the $B$ constant to be zero. As described above in Section II.B, for a low number fraction of the immiscible species (here the number fraction $x \sim$ 0.014) $B$ should be very close to zero. At 5 Mbar pressure we find $A = 3.65$ eV. The phase diagram was shown in Figure 4.

There have been at least two prior calculations concerning the phase separation of carbon and oxygen from liquid metallic oxygen. In a fashion similar to prior helium separation calculations, the assumption is made that the carbon or oxygen atoms are fully ionized. However, one investigation (Stevenson 1976) only calculated the critical temperatures and compositions for these mixtures. No phase diagrams were calculated. The critical compositions were found to be 0.086 and 0.064 for carbon and oxygen, respectively, which are ~ 100 times greater than one would expect to find in Jupiter or Saturn. It is unclear at what temperatures such low concentrations of carbon and oxygen would separate out. Brami, *et al.* (1979) calculated phase diagram for fully ionized carbon/hydrogen and oxygen/hydrogen mixtures, but only for pressures in excess of 3 Gbar. Their calculations at least qualitatively seem to indicate that very low concentrations of carbon or oxygen could become immiscible at temperatures on the order of 1-10 x $10^4$ K. The applicability of these calculations to the deep interiors of giant planets is not clear, since carbon and oxygen are not expected to be fully ionized.

In our "Ice" trial, Saturn has an initial olivine core of 0.14 $M_{Sat}$ and uniform heavy element distribution of $Z_{ice} = 0.145$ in the H/He envelope. When Saturn reaches $T_{eff} = 125$ K, the ices begin to separate out, and by the time the planet reaches $T_{eff} = 95$ K, the mass fraction $Z_{ice}$ of heavy elements in the H/He layer has fallen to 0.045. The entropy of the hydrogen-rich envelope increases slightly, since its mean molecular weight decreases, while the heavy-element core (with specific entropy set equal to zero) grows in size. Energy is released as mass shells near the core are turned from high-entropy H/He layers to zero-entropy ice layers. Figure 14 shows the cooling of Saturn with CNO separation, compared with homogeneous evolution of Jupiter and Saturn.

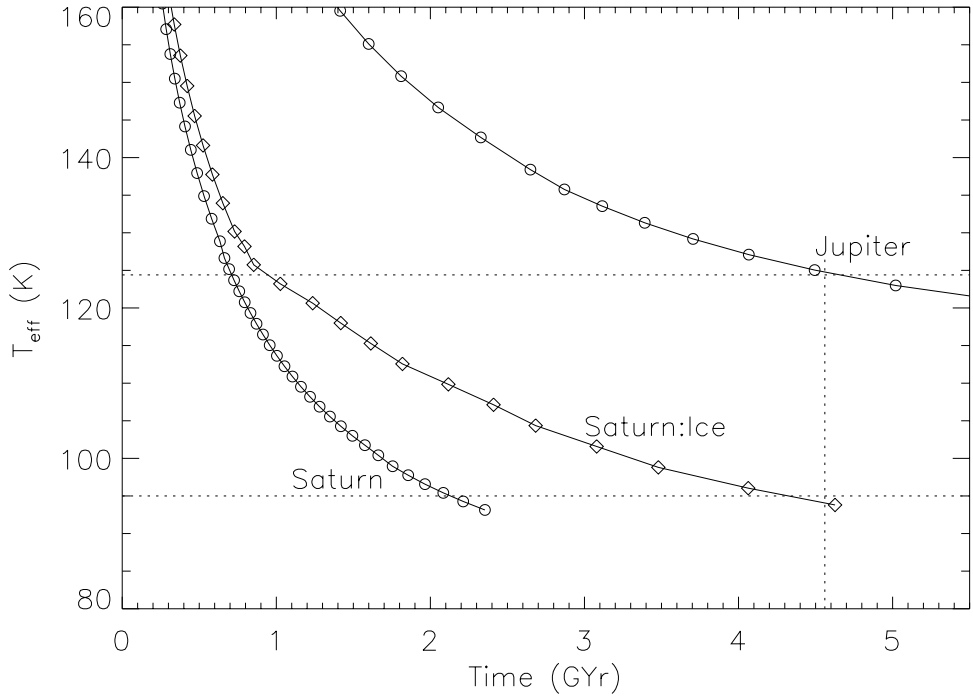

Figure 14. Evolution of Saturn with separation of heavy elements. Homogeneous evolutionary models are labeled "Saturn" and "Jupiter," while the evolution of Saturn with separation of CNO elements is labeled "Saturn:Ice." The phase diagram used is similar to that of Pfaffenzeller, et al (1995), with a positive $c_1$. Here the "ices" are lost from the entire hydrogen-rich envelope and form a layer of pure ice on the planet's core. Separation of CNO elements allows Saturn to reach $T_{eff} = 95.0$ K at an age close to the solar system's.

Although CNO separation starts at a planetary effective temperature greater than that for the onset of He separation, CNO separation cannot occur during the evolution of Jupiter for this phase diagram. Our best-fit Jupiter model has $Z_{ice}$ in the planet's envelope of 0.059. Jupiter's adiabat never drops to temperatures low enough to encounter an immiscibility region for $Z_{ice} = 0.059$.

Our CNO-separation evolutionary models can be seen as a proof of concept for heavy-element separation in Saturn. If the abundant CNO elements are initially enhanced by a factor of 5-10 in Saturn relative to solar abundance, the $A$ constant does not need to be unreasonably large for CNO separation to occur. An $A$ constant larger than that of helium is probably even likely, considering that insertion of a larger atom such as oxygen into liquid-metallic hydrogen should incur a larger enthalpy increase. Since Saturn is colder than Jupiter and is probably more enriched in heavy elements, a significant amount of heavy element separation (and extension of cooling) can occur in Saturn without any corresponding effect on Jupiter.

Determining the water abundance in Jupiter is important both for a census of oxygen in the solar system and because it has recently been put forward as a sensitive test for the formation of the planet. A model proposed by Owen, *et al.* (1999) predicts Jupiter's ice component came from ISM planetesimals that never were heated above 30 K. A prediction of this formation scenario is that Jupiter has three times solar abundance of water, the same as the enrichment that has already been detected for other elements. In the scenario of Gautier, *et al.* (2001a, 2001b) the planetesimals that gave rise to Jupiter's ice component were clathrate hydrates, and they predict Jupiter should have an oxygen abundance at least 9.4 times solar. Since we perform no detailed static models, and our equation of state for the ice component of the envelope of the planet (ANEOS $H_2O$ EOS) has been shown to be inaccurate at high pressure (Chau, *et al.* 2001), one should consider our quoted best-fit $Z_{ice}$ values to be schematic. The reader should be aware of the possibility that phase separation of water could lead to the planet's atmospheric abundance differing from that of the planet as a whole, although this scenario is probably more likely for Saturn than Jupiter.

### F. Summary of Trials

Table 4 summarizes the results for all of the phase diagrams explored in this paper. Figure 15 shows the $A$ and $B$ constants (at 5 Mbar) for all of the trials in Table 4. While these constants are not bounded by any fundamental constraints at present, the shaded area in Figure 15 shows the range of values for $A$ and $B$ that would be consistent with previous studies and with the theory presented in Section II. As we see, only two trials that give the correct Saturn age fall within this shaded region, Trial 8 and "Ice". These trials are marked in bold face in Table 4 and represent the preferred models of this paper. They also provide a prediction, both for possible future Saturn abundance measurements, and for experimental studies of mixtures of hydrogen with heavier elements at high pressure.

**Table 4**
**Saturn He Separation Models**

| Trial | H phase line | Phase diagram | $T_{eff}$ (K), onset | Final $Y_{atmos}$ | $Z_{ice}$ | $M_{core}$ ($M_{Sat}$) | Age (Gyr) | $A$ (eV) | $B$ |
|-------|------|------|------|------|------|------|------|------|------|
| 1 | PPT | Standard HDW | 107 | 0.215 | 0.04 | 0.20 | 2.9 | 1.3 | 0.69 |
| 2 | PPT | Modified HDW | 107 | 0.10 | 0.09 | 0.15 | 4.5 | 3.7 | 6.3 |
| 3 | PPT | Modified HDW | 98.5 | 0.13 | 0.08 | 0.16 | 4.6 | 4.6 | 8.7 |
| 4 | 1 Mbar | Modified HDW | 107 | 0.06 | 0.09 | 0.15 | 4.6 | 2.8 | 3.8 |
| 5 | 1 Mbar | Modified HDW | 98.5 | 0.09 | 0.08 | 0.16 | 4.5 | 3.4 | 5.5 |
| 6 | 3 Mbar | Modified HDW | 107 | 0.14 | 0.09 | 0.15 | 4.9 | 2.1 | 2.2 |
| 7 | 3 Mbar | Modified HDW | 98.5 | 0.16 | 0.08 | 0.16 | 4.6 | 3.4 | 5.5 |
| **8** | **N/A** | **Mod. Pfaf.** | **107** | **0.185** | **0.085** | **0.17** | **4.5** | **2.07** | **0.28** |
| 9 | N/A | Mod. Pfaf. | 98.5 | 0.20 | 0.065 | 0.185 | 4.4 | 3.2 | 2.1 |
| **Ice** | **N/A** | **Mod. Pfaf.** | **125** | **0.045*** | **0.145*** | **0.14** | **4.3** | **3.65** | **0** |

Preferred models in **boldface**. The ages for models 1-7 indicate Saturn's age for the maximum extension of cooling, as discussed in Section III.B.
*Final $Z_{ice}$; initial $Z_{ice}$

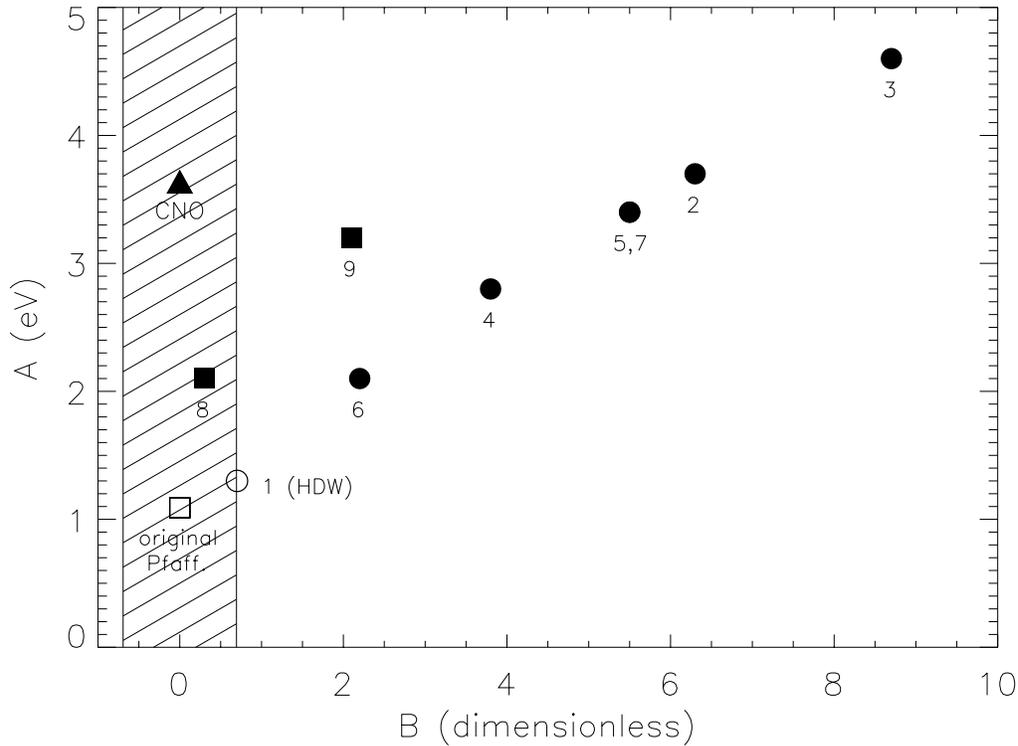

Figure 15. *A* and *B* parameters at 5 Mbar from Eq. (13) for Saturn models tabulated in Table 4. The open square and open circle are respectively the unmodified Pfaffenzeller *et al.* (1995) and HDW theories. Solid squares are modified Pfaffenzeller phase diagrams, while solid dots are modified HDW phase diagrams. The solid triangle corresponds to a phase diagram for separation of CNO heavy elements as discussed in Section III.E. The latter model can work with *B* = 0 because the mole fraction of CNO elements is much less than that of He (although still elevated with respect to solar mixing ratios). The hatched region shows models with plausible *B* values, based on prior calculated phase diagrams.

## IV. DISCUSSION

### A. Implications for Primordial Saturn Core Mass

Note the rocky (primordial) core mass for the Saturn models in Table 4. This mass can be as low as ~14-15 $M_\oplus$, due to the large redistribution of helium towards the center of the planet for certain extreme models. Although the models with smaller rock core masses have unphysical H/He phase diagrams, the ice-separation model is plausible and has a primordial core mass of ~14 $M_\oplus$. The point is that a smaller rocky core, together with either very helium-rich or ice-rich layers near the planet's core can reproduce Saturn's observed moment of inertia. Fourteen $M_\oplus$ is only 4 $M_\oplus$ larger than the mass derived for Jupiter's core, using the SCVH "interpolated" equation of state for hydrogen and helium. This would fit neatly in the core accretion theory for the formation of the giant planets, since it would then seem both planets began to accrete gases onto their cores when they reached nearly the same "critical mass." But as noted below, there may be other complications as to how the primordial core mass relates to the current core mass.

### C. Speeding Jupiter's Evolution

We have explored the evolution of Saturn and shown that various binary phase diagrams allow Saturn to reach the age of the solar system at $T_{eff} = 95.0$ K. All require substantial redistribution of helium or other abundant elements in the planet's interior, leading to a corresponding decrease in the mass fraction of that element in the planet's visible atmosphere. In our current theoretical understanding, any such separation in Jupiter would lead to a prolongation of evolution, worsening the agreement with the age of the solar system attained with homogeneous models. The observation that Jupiter's atmospheric helium mass fraction is 0.231, less than the protosolar value of 0.27, remains problematic and significant. Surely helium separation is a likely explanation, and this case is bolstered by the observation that neon is depleted in Jupiter's atmosphere (see Table 2), which Roulston and Stevenson (1995) attributed to the incorporation of neon into helium droplets separating out from the hydrogen. A recent review (Hubbard *et al.* 2002) points out that reduction of the neon concentration by about a factor of ten, accompanied by the indicated reduction of the helium concentration, would require an unusually high solubility of neon atoms in the helium droplets. But if this is the case, some other factor must be involved that without helium separation would allow the planet to cool more quickly than current homogeneous models predict.

There are several candidates for a process that could lead to faster cooling to counteract the prolongation of cooling due to phase separation. A full range of possibilities is discussed in Guillot *et al.* (2002). Before any additional processes are invoked, perhaps the first item to consider is the new generation of model atmospheres that have been developed in the past few years for EGPs and BDs. They incorporate better knowledge of chemistry, cloud formation, and stellar irradiation, and will lead to a more accurate grid of model atmospheres for Jupiter and Saturn. Work on developing a new grid suitable for evolution calculations is currently in progress. This grid will incorporate the planets' metal enhancement (relative to solar) and the effect of absorption of solar photons in thin outer atmosphere, which tends to make Jupiter and Saturn's atmosphere slightly more isothermal than isolated atmosphere models predict. It seems likely that these improvements will lead to slightly larger cooling ages for both planets. In concert with better atmospheric models, progress has been made on the albedos of giant planets, which will allow a more refined estimate of the amount of absorbed solar flux over the planet's history (Sudarsky *et al*. 2000).

The equation of state in the region near 1-5 Mbar is still uncertain, as discussed in detail earlier. This has a large effect on evolutionary models of Jupiter and Saturn. For example, Guillot (1999) calculated an age of 2.0 Gyr for an adiabatic Saturn using the SCVH "interpolated" EOS, while using the SCVH "PPT" EOS gave 2.7 Gyr. The uncertainty in the EOS alone probably adds an uncertainty of $\pm 0.005$ to 0.01 in the final calculated $Y_{atmos}$ for Saturn evolutionary models. Slightly more or less He may need to separate out in order to give the correct extension of cooling. Perhaps differences between calculations and experiment will narrow in the next few years. Further experiments pushing to higher pressures may help to elucidate the possibility of a zone that is unconditionally stable to convection, as discussed in Section I.C. Such a zone might offer a possibility for modest diffusive separation of elements. Likewise, experimental measurements of the high-pressure properties of H-He mixtures may provide information about the value of $A$ and the sign of its pressure dependence.

An intriguing possibility for accelerating Jupiter's (and Saturn's) evolution could be core erosion by convective plumes. If heavy elements were eroded throughout the life of the planet and distributed in the interior against the force of gravity, some fraction of the planet's internal energy could be transformed into gravitational potential energy. Less of the planet's internal energy would have to be radiated to space, leading to faster evolution. This hypothesis would also be a natural explanation for the origin of the heavy element enhancement in the envelopes of the two planets (Guillot *et al.* 2002). A possible next step for inhomogeneous evolutionary models of Jupiter and Saturn would be to try to create a physically-reasonable phase diagram such that Jupiter enters the region of He immiscibility and its $Y_{atmos}$ drops to 0.231 at 4.56 Gyr, while Saturn is supplied sufficient energy via He separation to reach 4.56 Gyr. Some other process, such as core erosion and redistribution may have to be invoked to counteract the cooling extension of Jupiter. A study of that nature will be undertaken as a revised model atmosphere grid and high-pressure equation of state become available.

## V. CONCLUSIONS

If the recalculated Conrath and Gautier (2000) helium abundance for Saturn ($Y_{atmos} = 0.215 \pm 0.035$) is correct, a few statements can be made about the phase diagram of hydrogen and helium. If helium separation is the planet's only additional energy source, models in which helium rains out and settles onto the planet's core, rather than remixing at deeper levels, are preferred. Thus, the phase diagrams of Stevenson (1975) and HDW, which predict a narrow region of helium separation, and modest extension of Saturn's cooling (less than 1 Gyr) are likely to be inapplicable to Jupiter and Saturn. Models where helium is free to rain down to the planet's core lead to a large energy release with only moderate depletion of helium in the planet's atmosphere. As we show, this could be accomplished by a phase diagram similar to that of Pfaffenzeller *et al.* (1995), but with about twice as large an $A$, in which the immiscibility region runs roughly parallel to the planet's adiabats and maintains this shape past a pressure of ~ 10 Mbar. With this phase diagram Saturn's cooling can be extended so the planet reaches its current effective temperature at the age of the solar system, while the He mass fraction of its atmosphere falls from a protosolar $Y$ of 0.27 to 0.185. Separation of heavy and abundant CNO elements could affect the evolution of both planets, but this conclusion is necessarily tentative.

New observations from the Cassini spacecraft that could lead to another measurement of the helium abundance in Saturn's atmosphere will be extremely valuable. Even if higher precision than that obtained by Conrath and Gautier (2000) cannot be achieved, a new determination in support of their work would lend confidence that we are finally closing in on the correct value of Saturn's atmospheric helium abundance.

## ACKNOWLEDGMENTS
We thank the referees, Tristan Guillot and Barney Conrath, for helpful comments, and Daniel Gautier, Jonathan Lunine, Adam Burrows, Jason Barnes, Curtis Cooper, and Rip Collins for interesting conversations.

This research was supported by Grants NAG5-10760 (NASA Astrophysics Theory Program), NAG5-10629 (NASA Origins of Solar Systems Program), and NAG5-8906 (NASA Planetary Geology and Geophysics Program).

## REFERENCES

Allende Prieto, C., D. L. Lambert, and M. Asplund 2001. The forbidden abundance of oxygen in the Sun. *Astrophys. J.* **556**, L63-L66.

Bahcall, J. N., M. H. Pinsonneault, and G. J. Wasserburg 1995. Solar models with helium and heavy-element diffusion. *Reviews of Modern Physics* **67**, 781-808.

Brami, B., J.P. Hansen, F. Joly 1979. Phase separation of highly dissymmetric binary ionic mixtures. *Physica* A **95**, 505-525.

Burrows, A., M. Marley, W. B. Hubbard, J. I. Lunine, T. Guillot, D. Saumon, R. Freedman, D. Sudarsky, and C. Sharp 1997. A nongray theory of extrasolar giant planets and brown dwarfs. *Astrophys. J.* **491**, 856-875.

Burrows, A., W. B. Hubbard, J. I. Lunine, and J. Liebert 2001. The theory of brown dwarfs and extrasolar giant planets. *Reviews of Modern Physics* **73**, 719-765.

Chau, R., Mitchell, A.C., Minich, R.W., Nellis, W.J. 2001. Electrical conductivity of water compressed dynamically to pressures of 70-180 GPa (0.7-1.8 Mbar). *Journal of Chem. Phys.* **114**, 1361-1365.

Collins, G. W., L. B. Da Silva, P. Celliers, D. M. Gold, M. E. Foord, R. J. Wallace, A. Ng, S. V. Weber, K. S. Budil, and R. Cauble 1998. Measurements of the equation of state of deuterium at the fluid insulator-metal transition. *Science* **281**, 1178-1181.

Conrath, B. J., R. A. Hanel, and R. E. Samuelson 1989. Thermal structure and heat balance of the outer planets. In *Origin and Evolution of Planetary and Satellite Atmospheres* (S. K. Atreya, J. B. Pollack, and M. S. Matthews, Eds.), pp. 513-538. Univ. of Arizona Press, Tucson.

Conrath, B. J., D. Gautier, R. A. Hanel, and J. S. Hornstein 1984. The helium abundance of Saturn from Voyager measurements. *Astrophysical J.* **282**, 807-815.

Conrath, B. J., and D. Gautier 2000. Saturn helium abundance: a reanalysis of Voyager measurements. *Icarus* **144**, 124-134.

Gautier, D., B. Conrath, M. Flasar, R. Hanel, V. Kunde, A. Chedin, and N. Scott 1981. The helium abundance of Jupiter from Voyager. *J. Geophys. Res.* **86**, 8713-8720.

Gautier, D., F. Hersant, O. Mousis, J.I. Lunine 2001a Enrichments in Volatiles in Jupiter: A New Interpretation of the Galileo Measurements. *Astrophys. J.* **550**, L227-L230.

Gautier, D., F. Hersant, O. Mousis, J.I. Lunine 2001b Erratum: Enrichments in Volatiles in Jupiter: A New Interpretation of the Galileo Measurements. *Astrophys. J.* **559**, L183-L183.

Grevesse, N., and A.J. Sauval 1998. Standard solar composition. *Space Science Reviews* **85**, 161-174.

Grossman, A. S., J. B. Pollack, R. T. Reynolds, A. L. Summers, and H. C. Graboske 1980. The effect of dense cores on the structure and evolution of Jupiter and Saturn. *Icarus* **42**, 358-379.

Guillot, T. 1999. A comparison of the interiors of Jupiter and Saturn. *Planetary and Space Science* **47**, 1183-1200.

Guillot, T., D. Gautier, G. Chabrier, and B. Mosser 1994. Are the giant planets fully convective? *Icarus* **112**, 337-353.

Guillot, T., D. Gautier, and W.B. Hubbard 1997. New constraints on the composition of Jupiter from Galileo measurements and interior models. *Icarus* **130**, 534-539.

Guillot, T., D. J. Stevenson, W. B. Hubbard, and D. Saumon 2002. The interior of Jupiter. In *Jupiter - The Planet, Satellites and Magnetosphere* (F. Bagenal, Ed.), submitted.

Hubbard, W. B. 1970. Structure of Jupiter: chemical composition, contraction, and rotation. *Astrophys. J.* **162**, 687-697.

Hubbard, W. B. 1977. The Jovian surface condition and cooling rate. *Icarus* **30**, 305-310.

Hubbard, W. B., and D.J. Stevenson 1984. Interior Structure of Saturn. In *Saturn* (T. Gehrels and M. Matthews, Eds.), pp. 47-87. Univ. of Arizona Press, Tucson.

Hubbard, W. B., and H. E. DeWitt 1985. Statistical mechanics of light elements at high pressure. VII. A perturbative free energy for arbitrary mixtures of H and He. *Astrophys. J.* **290**, 388-393.

Hubbard, W. B., T. Guillot, M. S. Marley, A. Burrows, J.I. Lunine, and D. Saumon 1999. Comparative Evolution of Jupiter and Saturn. *Planetary and Space Science* **47**, 1175-1182.

Hubbard, W. B., A. Burrows, and J.I. Lunine 2002. The theory of giant planets. *Annual Review of Astronomy and Astrophysics* **40**, 103-136.

Knudson, M.D., D. L. Hanson, J. E. Bailey, C. A. Hall, J. R. Asay, W. W. Anderson 2001. Equation of State Measurements in Liquid Deuterium to 70 GPa. *Physical Review Letters* **87**, 225501-1.

Niemann, H. B., J.A. Haberman, W.T. Kasprzak, R.F. Beebe, and L.F. Huber, GP-J-NMS-3-ENTRY-V1.0, NASA Planetary Data System.

Owen, T., P. Mahaffy, H.B. Niemann, S. Atreya, T. Donahue, A. Bar-Nun, I. de Pater 1999. A low-temperature origin for the planetesimals that formed Jupiter. *Nature* **402**, 269-270.

Pfaffenzeller, O., D. Hohl, and P. Ballone 1995. Miscibility of hydrogen and helium under astrophysical conditions. *Phys. Rev. Lett.* **74**, 2599-2602.

Pollack, J. B., A. S. Grossman, R. Moore, and H. C. Graboske, Jr. 1977. A calculation of Saturn's gravitational contraction history. *Icarus* **30**, 111-128.

Ross M. 1998. Linear-mixing model for shock-compressed liquid deuterium. *Phys. Rev. B* **58**,669-677.

Roulston, M.S., and D.J. Stevenson 1995. Prediction of neon depletion in Jupiter's atmosphere. *EOS* **76**, 343 (abstract).

Saumon, D, G. Chabrier, and H. M. van Horn 1995. An equation of state for low-mass stars and giant planets. *Astrophys. J. Suppl.* **99**, 713-741.

Showman, A. P., and A. P. Ingersoll. Interpretation of Galileo probe data and implications for Jupiter's dry downdrafts. *Icarus* **132**, 205-220.

Stevenson, D. J. 1975. Miscibility gaps in fully pressure-ionized binary alloys. *Phys. Letters A* **58**, 282-284.

Stevenson, D. J. 1976. Thermodynamics and phase separation of dense fully ionized hydrogen-helium fluid mixtures. *Phys. Rev. B* **12**, 3999-4007.

Stevenson, D. J. 1979. Solubility of helium in metallic hydrogen. *J. Phys. F: Metal Phys.* **9**, 791-801.

Stevenson, D. J., and E. E. Salpeter 1977a. The phase diagram and transport properties for hydrogen-helium fluid planets. *Astrophys. J. Suppl.* **35**,221-237.

Stevenson, D. J., and E. E. Salpeter 1977b. The dynamics and helium distribution in hydrogen-helium fluid planets. *Astrophys. J. Suppl.* **35,**239-261.

Sudarsky, D., A. Burrows, and P. Pinto 2000. Albedo and reflection spectra of extrasolar giant planets. *Astrophys. J.* **538**, 885-903.

Von Zahn, U., D.M. Hunten, R.F. Beebe, and L.F. Huber 2000. GP-J-HAD-3-ENTRY-V1.0, NASA Planetary Data System.